\documentclass[format=acmsmall, review=false, screen=true]{acmart}
\usepackage{svg}
\usepackage{graphicx}
\usepackage{array}
\usepackage{url}
\usepackage{fancybox}
\usepackage{multirow}
\usepackage{amssymb}
\usepackage{color}

\usepackage{colortbl}
\usepackage{balance}
\usepackage{fancyhdr}

\usepackage{subfig}
\usepackage{diagbox}
\usepackage{bbding}
\usepackage{placeins}

\usepackage{booktabs}
\usepackage{enumitem}
\usepackage{xcolor,lipsum}

\definecolor{light-gray}{rgb}{.906,  .902,  .902}

\newcommand{\rev}[1]{\textcolor{black}{#1}}
\newcommand{\revv}[1]{\textcolor{black}{#1}}

\acmJournal{TOSEM}
\acmVolume{9}
\acmNumber{4}
\acmArticle{39}
\acmYear{2019}
\acmMonth{3}
\copyrightyear{2009}
\acmArticleSeq{9}

\setcopyright{acmlicensed}

\acmDOI{0000001.0000001}

\received{Mar 2019}
\received[revised]{?? 2019}
\received[accepted]{?? 2019}

\usepackage{bm}

\begin{document}
\title{Generating Question Titles for Stack Overflow from Mined Code Snippets}
\titlenote{Corresponding Authors: Xin Xia}

\author{Zhipeng GAO}
\affiliation{%
  \institution{Monash University}
  \city{Melbourne,}
  \state{VIC}
  \postcode{3168}
  \country{Australia}
  }
\email{zhipeng.gao@monash.edu}

\author{Xin Xia}
\affiliation{%
  \institution{Monash University}
  \city{Melbourne,}
  \state{VIC}
  \postcode{3168}
  \country{Australia}
  }
\email{xin.xia@monash.edu}

\author{John Grundy}
\affiliation{%
  \institution{Monash University}
  \city{Melbourne,}
  \state{VIC}
  \postcode{3168}
  \country{Australia}
  }
\email{john.grundy@monash.edu}

\author{David Lo}
\affiliation{%
  \institution{Singapore Management University}
  \city{Singapore,}
    \country{Singapore}
  }
\email{davidlo@smu.edu.sg}

\author{Yuan-Fang Li}
\affiliation{%
  \institution{Monash University}
  \city{Melbourne,}
  \state{VIC}
  \postcode{3168}
  \country{Australia}
  }
\email{yuanfang.li@monash.edu}

\begin{abstract}
Stack Overflow has been heavily used by software developers as a popular way to seek programming-related information from peers via the internet. The Stack Overflow community recommends users to provide the related code snippet when they are creating a question to help others better understand it and offer their help. 
\rev{
\revv{Previous studies have shown that} a significant number of these questions are of low-quality and not attractive to other potential experts in Stack Overflow. These poorly asked questions are less likely to receive useful answers and hinder the overall knowledge generation and sharing process. Considering one of the reasons for introducing low-quality questions in SO is that many developers may not be able to clarify and summarize the key problems behind their presented code snippets due to their lack of knowledge and terminology related to the problem, and/or their poor writing skills, in this study  we propose an approach to assist developers in writing high-quality questions by automatically generating question titles for a code snippet using a  deep sequence-to-sequence learning approach.} 
Our approach is fully data-driven and uses an \textit{attention} mechanism to perform better content selection, a \textit{copy} mechanism to handle the rare-words problem and a \textit{coverage} mechanism to eliminate word repetition problem. 
\rev{
We evaluate our approach on Stack Overflow datasets  over a variety of programming languages (e.g., Python, Java, Javascript, C\# and SQL) and our experimental results show that our approach significantly outperforms several state-of-the-art baselines in both automatic and human evaluation. We have released our code and datasets to facilitate other researchers to verify their ideas and inspire the follow up work.
}

\end{abstract}

%
%

%
%


\maketitle

\renewcommand{\shortauthors}{Zhipeng GAO et al.}

\section{Introduction}
\label{sec:intro}

In recent years, question and answer (Q\&A) platforms have become one of the most important user generated content (UGC) portals.
Compared with general Q\&A sites such as Quora\footnote{https://www.quora.com/} and Yahoo! Answers\footnote{https://answers.yahoo.com/}, Stack Overflow\footnote{https://stackoverflow.com/} is a vertical domain Q\&A site, its content covers the specific domain of computer science and programming.
Q\&A sites, such as Stack Overflow, are quite open and have little restrictions, which allow their users to post their problems in detail. Most of the questions will be answered by users who are often domain experts.

Stack Overflow (SO) has been used by developers as one of the most common ways to seek coding and related information on the web. Millions of developers now use Stack Overflow to search for high-quality questions to their programming problems, and Stack Overflow has also become a knowledge base for people to learn programming skills by browsing high-quality questions and answers.
The success of Stack Overflow and of community-based question and answer sites in general depends heavily on the will of the users to answer others' questions. Intuitively, an effectively written question can increase the chance of getting help. This is beneficial not only for the information seekers, since it increases the likelihood of receiving support, but also for the whole community as well, since it enhances the behavior of effective knowledge sharing. A high-quality question is likely to obtain more attention from potential answerers. On the other hand, low-quality questions may discourage potential helpers ~\cite{mamykina2011design, calefato2018ask, nie2017data, anderson2012discovering, yang2014asking, jin2019edits}.


\rev{
To help users effectively write questions, Stack Overflow has developed a list of quality assurance guidelines\footnote{https://stackoverflow.com/help/how-to-ask} for community members. However, despite the detailed guidelines, a significant number of questions submitted to SO are of low-quality ~\cite{correa2014chaff, arora2015good}. Previous research has provided some insight into the analysis of question quality on Stack Overflow~\cite{correa2013fit, correa2014chaff, arora2015good, anderson2012discovering, li2012analyzing, yao2013want, liu2013question, zhang2018code, duijn2015quality, Trienes2019IdentifyingUQ}. Correa and Sureka~\cite{correa2014chaff} investigated closed questions on SO, which suggest that the good question should contain enough code for others to reproduce the problem.
Arora et al.~\cite{arora2015good} proposed a novel method for improving the question quality prediction accuracy by making use of content extracted from previously asked similar questions in the forum. More recent work~\cite{Trienes2019IdentifyingUQ} studied the way of identifying unclear questions in CQA websites. However, all of the work focuses on predicting the poor quality questions and how to increase the accuracy of the predictions, more in-depth research of dealing with the low-quality questions is still lacking.  
To the best of our knowledge, this is the first work that investigates the possibility of automatically improving low-quality questions in Stack Overflow. 
}
\rev{\revv{Previous studies~\cite{correa2013fit, Trienes2019IdentifyingUQ, Toth2019TowardsAA} have shown that one of the major reasons for the introduction of low-quality questions is that developers do not create \textit{informative} question titles.} Considering information seekers may lack the knowledge and terminology related to their questions and/or their writing may be poor, formulating a clear question title and questioning on the key problems could be a non-trivial task for some developers. Lacking important terminology and pool expression may happen even more often when the developer is less experienced or less proficient in English.
}

\rev{
Among the Stack Overflow quality assurance guidelines, one of which is that developers should attach code snippets to questions for the sake of clarity and completeness of information, which lead to an impressive number of code snippets together with relevant natural language descriptions accumulated in Stack Overflow over the years. 
}
Some prior work has investigated retrieving or generating code snippets based on natural language queries, as well as annotating code snippets using natural language (e.g., ~\cite{franks2015cacheca, allamanis2015bimodal, giordani2009semantic, iyer2016summarizing, keivanloo2014spotting, ling2016latent, oda2015learning, vinayakarao2017anne, desai2016program, locascio2016neural, yin2017syntactic, wong2013autocomment, giordani2012translating, li2014nalir, gulwani2014nlyze, hu2018deep}). 
However, to the best of our knowledge, there have been no studies dedicated to the question generation\footnote{``question generation'' in this paper is to generate the question titles for a Stack Overflow post.} task in Stack Overflow, especially generating questions based on a code snippet. 

\begin{figure}
\centerline{\includegraphics[width=0.5\textwidth]{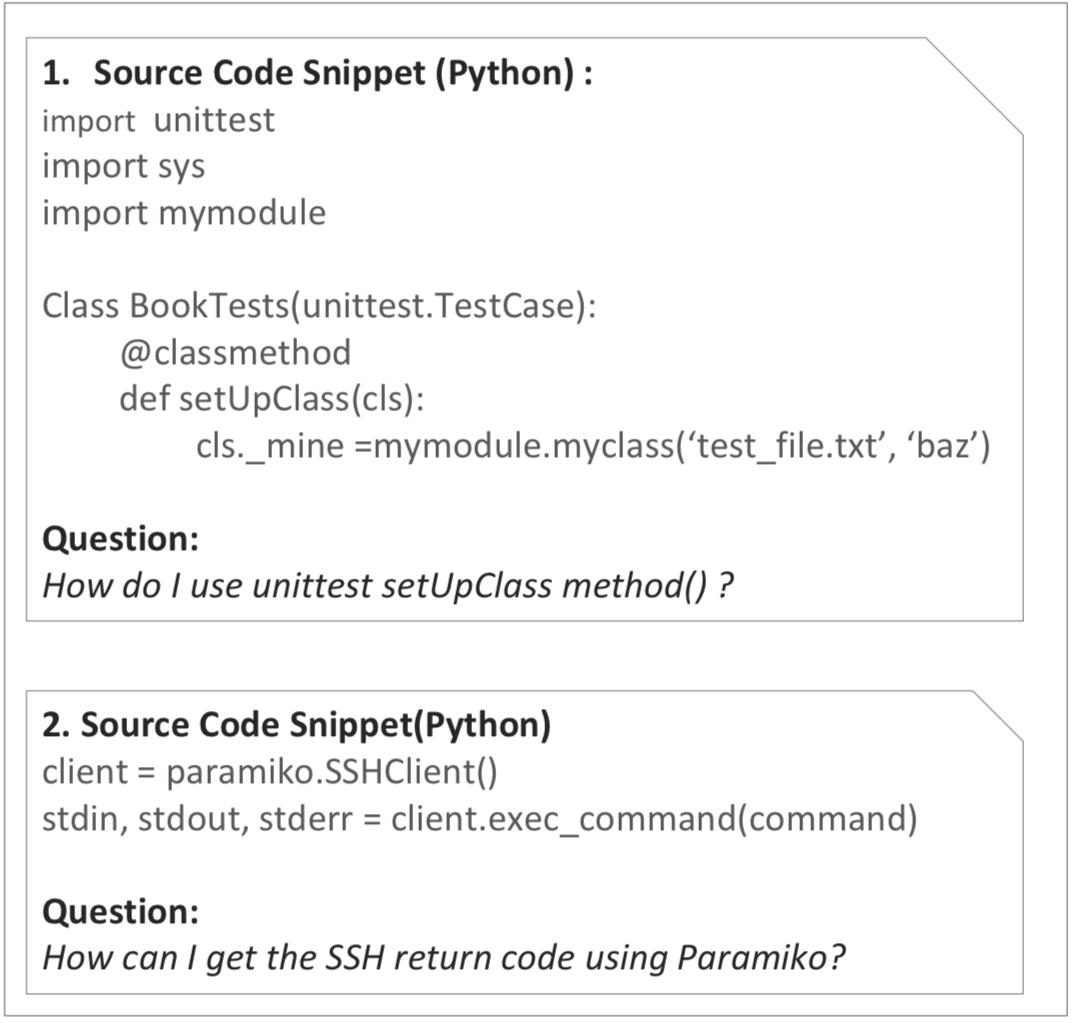}}
\vspace*{-0pt}
\caption{Example Code Snippet \& Question Pairs}
\label{fig:example}\vspace{-0.0cm}
\end{figure}

Fig.~\ref{fig:example} shows some example code snippets and corresponding question \rev{titles} in Stack Overflow. Generating such a question \rev{title} is often a challenging task since the corpus not only includes natural language text, but also complex code text. Moreover, some rare tokens occur among the code snippet, such as ``setUpClass'' and ``Paramiko'' illustrated in the aforementioned examples.

We propose an approach to help developers write high-quality questions based on their code snippets by automatically generating question titles from given code snippets. We frame this question generation task in Stack Overflow as a sequence-to-sequence learning problem, which directly maps a code snippet to a question.
To solve this novel task, we propose an end-to-end sequence-to-sequence system, enhanced with an \textit{attention} mechanism ~\cite{bahdanau2014neural} to perform better content selection, a \textit{copy} mechanism ~\cite{gu2016incorporating} to handle the rare-words problem, as well as a \textit{coverage} mechanism~\cite{tu2016modeling} to avoid meaningless repetition. Our system consists of two components: a source-code encoder and a question decoder. Particularly, the code snippet is transformed by a source-code encoder into a vector representation. When it comes to the decoding process, the question decoder reads the code embeddings to generate the target question \rev{titles}. Moreover, our approach is fully data-driven and does not rely on hand-crafted rules.

To demonstrate the effectiveness of our model, we evaluated it using automatic metrics such as BLEU~\cite{papineni2002bleu} and ROUGE~\cite{lin2004rouge} score, together with a human evaluation for naturalness and relevance of the output. 
\rev{We also performed a practical manual evaluation to measure the effectiveness of our approach for improving the low-quality questions in Stack Overflow.}
From the automatic evaluation, we found that our approach significantly outperforms a collection of state-of-the-art baselines, including the approach based on information retrieval~\cite{robertson1994some}, a statistical machine translation approach~\cite{koehn2007moses}, and an existing sequence-to-sequence architecture approach in commit message generation~\cite{jiang2017automatically}. For human evaluation, questions generated by our system are also rated as more natural and relevant to the code snippet compared with the baselines. \rev{The practical manual evaluation shows that our approach can improve the low-quality question titles in terms of Clearness, Fitness and Willingness.}  

In summary, this paper makes the following three main contributions:
\begin{itemize}
    \item We propose a novel question generation task based on a sequence-to-sequence learning approach, which can help developers to phrase \rev{high-quality question titles from given code snippets}. Enhanced with the \textit{attention} mechanism, our model can perform the better content selection, with the help of and \textit{copy} mechanism and \textit{coverage} mechanism, our model can manage rare word in the input corpus and avoid the meaningless repetitions. \rev{To the best of our knowledge, this is the first work which investigates the possibility of improving the low-quality questions in Stack Overflow.}
    \item We performed comprehensive evaluations on  Stack Overflow datasets to demonstrate the effectiveness and superiority of our approach. Our system outperforms strong baselines by a large margin and achieves state of the art performance.
    
    \item \rev{
    We collected more than 1M $\langle$\textit{code snippet, \textit{question}}$\rangle$ pairs from Stack Overflow, which covers a variety of programming languages (e.g., Python, Java, Javascript, C\# and SQL). 
    We have released our code\footnote{\url{https://github.com/beyondacm/Code2Que}} and datasets~\cite{zhipeng_gao_2020_3816592}
    to facilitate other researchers to repeat our work and verify their ideas. We also implemented a web service tool, named {\sc Code2Que}
    to facilitate developers and inspire the follow-up work.
    }
\end{itemize}

The rest of the paper is organized as follows.
\rev{
\revv{
Section~\ref{sec:related} presents key related work on question generation and relevant techniques.
Section~\ref{sec:pre} presents the motivation of this study.
}
Section~\ref{sec:approach} presents the details of our approach for the question generation task in Stack Overflow.
}
\rev{
Section~\ref{sec:eval} presents the experimental setup, the baseline methods and the evaluation metrics used in our study.
Section~\ref{sec:results} presents the detailed research questions and the evaluation results under each research question.
\revv{
Section~\ref{sec:discussion} presents the contribution of the paper and discusses the strength and weakness of this study.
Section~\ref{sec:threats} presents threats to validity of our approach.
}}
Section~\ref{sec:con} concludes the paper with possible future work. 

\section{Related Work}
\label{sec:related}
Due to the great value of Stack Overflow in helping software developers, there is a growing body of research conducted on Stack Overflow and its data. This section discusses various work in the literature closely related to our work, i.e., deep source code summarization, the empirical study of Stack Overflow on quality assurance, and different tasks by mining the Stack Overflow dataset. It is by no means a complete list of all relevant papers.

\subsection{Deep Source Code Summarization}
A number of previous works have proposed methods for mining the $\langle$natural language, code snippet$\rangle$ pairs, these techniques can be applied to tasks such as code summarization as well as commit message generation. (e.g., \cite{iyer2016summarizing}, \cite{hu2018deep}, \cite{jiang2017automatically},
\cite{wan2018improving}).

One similar work with ours is Iyer et at.\cite{iyer2016summarizing}. They proposed Code-NN, which uses an attentional sequence-to-sequence algorithm to summarize code snippets. This work is similar to our approach because our approach also uses an sequence-to-sequence model. However, there are three key differences between our approach and Code-NN. First, the goal of of Code-NN is summarizing source code snippets while the goal of our approach is generating questions from code snippets. Second, the Code-NN only incorporates attention mechanism while our approach also employs copy mechanism and coverage mechanism, which is more suitable for the specific task of question generation. Third, Code-NN needs to parse the code into AST, while most code snippets in SO are not parsable (e.g., the example code in Fig.~\ref{fig:generatedexamples}). Followed by Iyer's work, Hu et al.~\cite{hu2018deep} proposed to use the neural machine translation model on the code summarization with the assistance of the structural information (i.e., the AST). And Wan et al.~\cite{wan2018improving} applied deep reinforcement learning (i.e., tree structure recurrent neural network) to improve the performance of code summarization. Their approach also use AST as the input. All of the aforementioned studies rely on the AST structure of the source code, and note that most of the code in Stack Overflow are not parsable. Thus, \revv{the AST-based approaches can not apply to our work.
}


\subsection{Question Quality Study on Stack Overflow}
The general consensus is that the quality of user-generated content is a key factor to attract users to visit knowledge-sharing websites. Many studies have investigated the content quality in Stack Overflow (e.g., ~\cite{nasehi2012makes, yao2013want, yang2014asking, ponzanelli2014understanding, correa2013fit, correa2014chaff, arora2015good, anderson2012discovering, li2012analyzing, liu2013question, zhang2018code, duijn2015quality, Trienes2019IdentifyingUQ}).

For example,
Nasehi et al. ~\cite{nasehi2012makes} manually performed a qualitative assessment to investigate the important features of precise code examples in answers of 163 SO posts.
Yao et at. ~\cite{yao2013want} investigated quality prediction of both Q\&As on SO. The output revealed that answer quality is strongly positively associated with that of its question.
Yang et al.~\cite{yang2014asking} found that the number of edits on a question is a very good indicator of question quality.
Ponzanelli ~\cite{ponzanelli2014understanding} developed an approach to do automatic categorization of questions based on their quality.
Correa et al. ~\cite{correa2013fit} studied the closed questions in Stack Overflow, finding that the occurrence of code fragments is significant.

\rev{
All of the above mentioned studies are either predicting quality of the post or increasing the accuracy of predictions.
Different from the existing research, our approach is related to improve the quality of the questions. To the best of our knowledge, this is the first work which investigates the possibility of improving the low quality questions using code snippets in Stack Overflow.
}

\subsection{\rev{Machine/Deep Learning on Software Engineering}}
\rev{
Recently, an interesting direction of software engineering is to use machine/deep learning for different tasks to improve software development. Such as code search (e.g., ~\cite{gu2018deep, li2019neural, husain2019codesearchnet, allamanis2015bimodal}), clone detection (e.g., ~\cite{wang2020detecting, gao2020checking, white2016deep, buch2019learning, gao2019smartembed}), program repair (e.g,. ~\cite{white2019sorting, mesbah2019deepdelta, vasic2019neural, chen2019sequencer}), document (such as API and questions/answers/tags) recommendation (e.g., ~\cite{gu2016deep, gu2017deepam, xia2013tag, wang2015tagcombine, wang2018entagrec++, zhu2015building, gkotsis2014s, xu2017answerbot, singh2016using}).
}

\rev{
For code search tasks, Gu et al.~\cite{gu2018deep} proposed a deep code search model which uses two deep neural networks to encode source code and natural language description into a vector representation and then uses a cosine similarity function to calculate their similarity.
Allamanis et al. ~\cite{allamanis2015bimodal} proposed a system that uses Stackoverflow data and web search logs to create models for retrieving C\# code snippets given natural language questions and vice versa. 
For clone detection tasks, white et al.~\cite{white2016deep} first proposed a deep learning-based clone detection method to identify code clones via extracting features from program tokens. For program repair tasks, White et al.~\cite{white2019sorting} propose an automatic program repair approach, DeepRepair, which leverages a deep learning model to identify the similarity between code snippets. 
For document recommendation tasks,
Xia et al.~\cite{xia2013tag} developed a tool, called TagCombine, an automatic tag recommendation method which analyzes objects in software information sites. 
Gkotsis et al. ~\cite{gkotsis2014s} developed a novel approach to search and suggest the best answers through utilizing textual features.
Gangul et al. ~\cite{ganguly2015partially} examined the retrieval of a set of documents, which are closely associated with a newly posted question. Chen et al.~\cite{chen2016learning} studied cross-lingual question retrieval to assist non-native speakers more easily to retrieve relevant questions.
}

Although the aforementioned studies have utilized \revv{machine/deep learning} for different software development activities, to our best knowledge, no one has yet considered the question generation task in Stack Overflow.
In contrast to all previous work, we propose a novel approach to generate a question by a given code snippet. Our work is first to tackle such a task for helping developers to generate a question when presenting a given code snippet.

\section{Motivation}
\label{sec:pre}
\revv{
In this section, we first summarise the problem and our solution in this study. Following that, we present some example user scenarios of employing our approach in the software development process. We then show some motivating examples from Stack Overflow of the sorts of problems our work addresses.}
\subsection{The Problem and Our Solution}
\revv{
Despite the detailed guidelines provided by the community, a very large number of questions in Stack Overflow are of low-quality~\cite{arora2015good, correa2014chaff}. These poorly asked questions are often ambiguous, vague, and/or incomplete, and hardly attract potential experts to provide answers, thus hindering the progress of knowledge generation and sharing. In order to improve question quality, we need to improve title, body and tags. In this work, we focus on improving titles.  
The motivation for our work is that improving low-quality question titles can potentially be helpful in increasing the likelihood of getting help for the information seekers, as well as reducing the manual effort for quality maintenance of the CQA community.
We propose a novel approach to assist developers in posting high-quality questions by generating question titles for a given code snippet. Our approach provides benefit for the following tasks: (i) \emph{Question Improvement}: many developers can not post clear and/or informative questions due to their lack of knowledge and terminology related to the problem, and/or their poor english writing skills. Our approach can generate high-quality question titles for helping developers to summarize the key problems behind their presented code snippet. (ii) \emph{Edits Assistance}: the SO community has employed a collaborative editing mechanism to maintain a satisfactory quality level for the post. However, the editing process may require several interactions between the asker and other community members, thus delaying the answering and even causing questions to sink in the list of open issues. Our approach can be used as an automatic edit assistance tool to improve the question formulation process and reduce the manual effort for quality maintenance. (iii) \emph{Code Embeddings}: Another byproduct of our approach is the code embeddings generated by our approach. In this study, we have collected more than 1M code snippets which covers various programming languages such as Java, Python, Javascript, C\#, etc. All the code snippets are embedded into a high-dimensional vector space by our approach. A variety of applications such as code search (e.g., ~\cite{gu2018deep, li2019neural, husain2019codesearchnet}) , summarization (e.g., ~\cite{iyer2016summarizing, hu2018deep, jiang2017automatically, wan2018improving}), retrieval (e.g., ~\cite{chen2016learning, allamanis2013and, xu2018domain}), and API recommendation (e.g., ~\cite{gu2016deep,  gu2017deepam}) can benefit from the code embeddings used in our study. 
}


\subsection{\rev{Illustrative User Scenarios}}
\rev{
We implement our model as a standalone web application tool, called {\sc Code2Que}. Developers can copy and paste their code snippet to our tool to generate a question title for the code snippet. \revv{Meanwhile}, by utilizing the vector representation of the code snippets, {\sc Code2Que} also retrieves a list of top related questions in Stack Overflow and recommends them to the developers. The usage scenarios of our proposed tool are as follows:
}

\rev{
\textbf{Without Tool}. Consider Bob who is a developer, who is learning a new development framework. He is also a non-native English speaker with poor English writing skills. Daily, Bob encounters various programming problems during development. He locates the code that is the root cause of the problem, but he cannot figure it out. Due to his lack of the knowledge and terminology of the development framework being used, he does not even know how to most effectively search for answers to the problem on the Internet. Therefore, he creates a question in Stack Overflow, provides his code snippet in the question body according to the Stack Overflow guidelines, and then tries his best to write a question title to summarize the problem. Unfortunately, his question title turns out to be very unclear and uninformative, and there are few users attracted by his question. Bob waits for a long time but does not get any help.
}

\begin{figure}\vspace{-0.0cm}
\centerline{\includegraphics[width=0.75\textwidth]{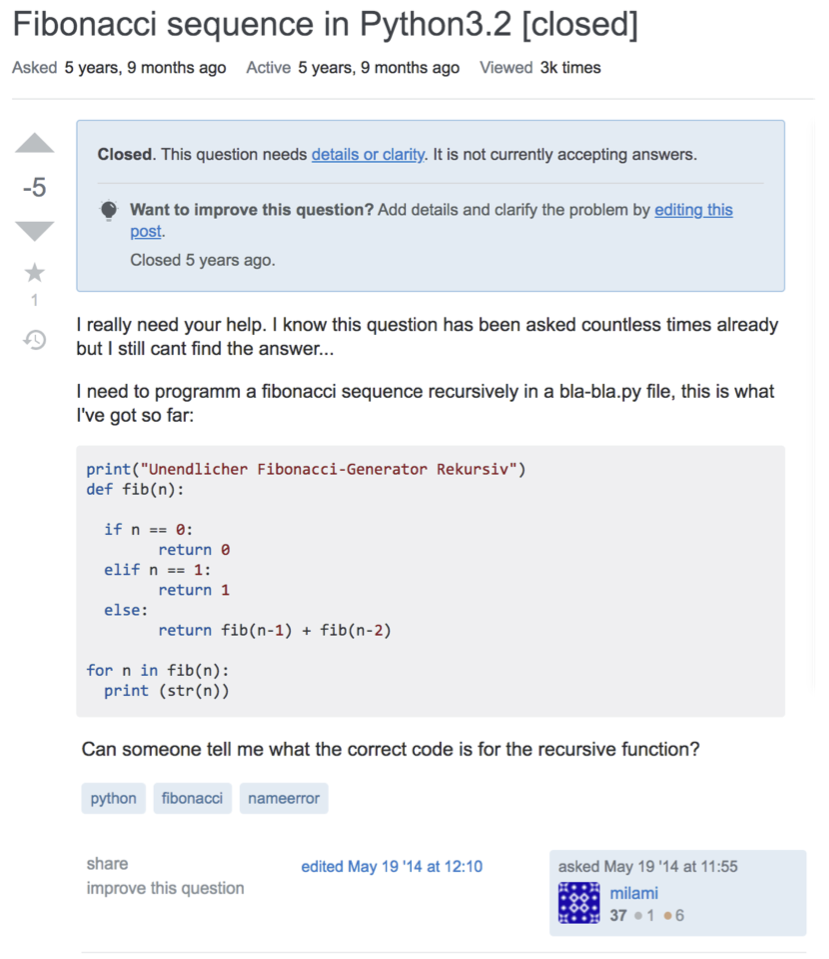}}
\vspace*{-0pt}
\caption{Example of Problem Questions Title (for Python)}
\label{fig:PythonQuestion}\vspace{0.0cm}
\end{figure}

\rev{
\textbf{With Tool}. Now consider that Bob adopts tool {\sc Code2Que}. Before he searches on the Internet, Bob copies his code snippet to our {\sc Code2Que} tool to generate a question title for the code snippet.
Bob uses the generated question as a query to search on the internet. The searching results are now closely related to the development framework, even though he is not very familiar with it. 
Bob can also quickly review a list of related questions in Stack Overflow which have a similar problem code snippet.
After going through these results, Bob can gain a better understanding of the problem that he is trying to solve and quickly fix the problem by himself.
Moreover, Bob can also go back to his earlier poorly asked questions, Bob can use our tool as an edit assistance tool on question titles for reformulating these low-quality questions.
Bob provides the code snippet in the question body and writes a question title based on the question title generated by our tool and the knowledge he learned from the results. This time, his question title is much more clear and informative and Bob's question soon attracts an expert of the development framework. With the help of this expert, Bob successfully figures his problem out.
}

\begin{figure}\vspace{-0.1cm}
\centerline{\includegraphics[width=0.75\textwidth]{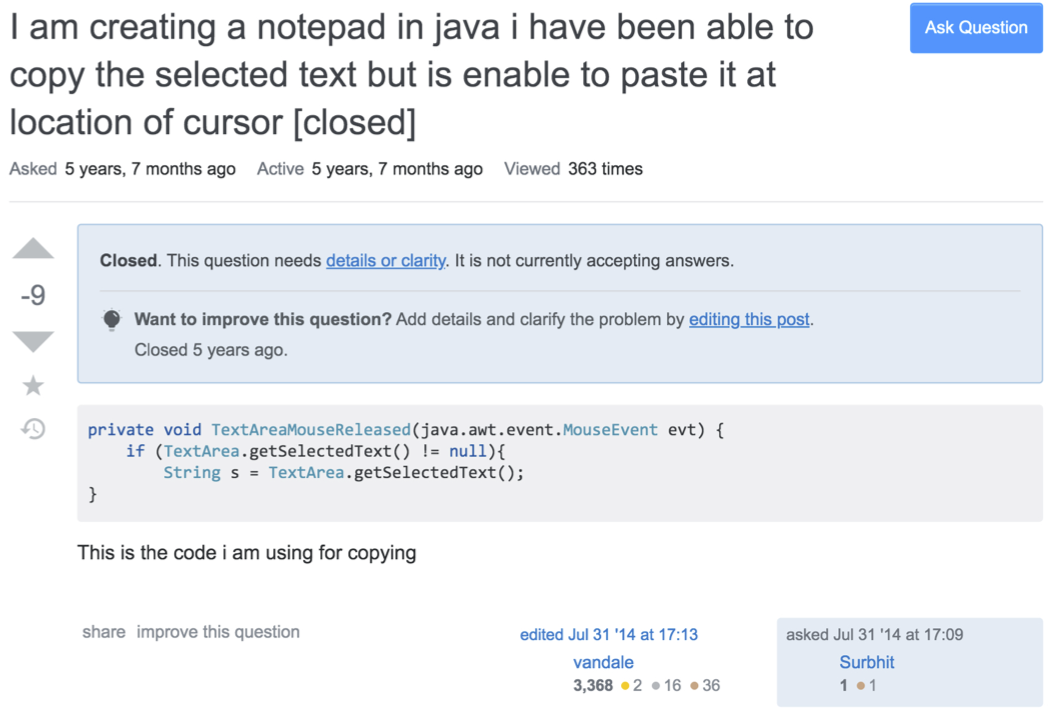}}
\vspace*{-0pt}
\caption{Example of Problem Questions Title (for Java)}
\label{fig:JavaQuestion}\vspace{0.1cm}
\end{figure}

\subsection{\revv{Motivating Examples}}

\rev{
A large number of questions have been closed by community members because their question titles are unclear and need further clarifying. For example, the screenshots in Fig.~\ref{fig:PythonQuestion} and Fig.~\ref{fig:JavaQuestion} show two examples of problematic Stack Overflow question titles.
Developers posted a question \emph{``Fibonacci sequence in Python3.2''} and \emph{``I am creating a notepad in java ... to paste it at location of cursor''} in Stack Overflow. They attached their code snippet and tried to explain the key meaning of their problems. However, such question titles are still very uninformative (in Fig.~\ref{fig:PythonQuestion}) and confusing (in Fig.~\ref{fig:JavaQuestion}). Both of these questions have been marked as having lack of clarity and need to be further improved upon.
Such titles run a real risk of not being found by the ideal people to answer them, may make potential question answering users lose interest, or make users who may answer them have to painstakingly browse the additional paragraph to understand the key point. All reduce the likelihood of them giving help.
}

\rev{
Using the tool {\sc Code2Que} described in this paper, we can provide a way to automate the process of improving such poor quality question titles, which is potentially helpful in reducing the manual effort for the quality maintenance of CQA forums. 
Based on the developer's code snippet, the generated question title by our tool is \emph{``how to find the fibonacci series through recursion?''} for the code snippet shown in Fig.~\ref{fig:PythonQuestion} and \emph{``how to change the string value in textarea field using java?''} for the code snippet shown in Fig.~\ref{fig:JavaQuestion}. These newly generated question titles are much more clear and informative to readers, and also questioning on the key problems of the user's concern. 
This is helpful for the potential helpers to understand the key problems of the question better and also for the askers to formulate a related question better.
}

\section{Approach}
\label{sec:approach}
\begin{figure*}\vspace{-0.0cm}
\centerline{\includegraphics[width=0.98\textwidth]{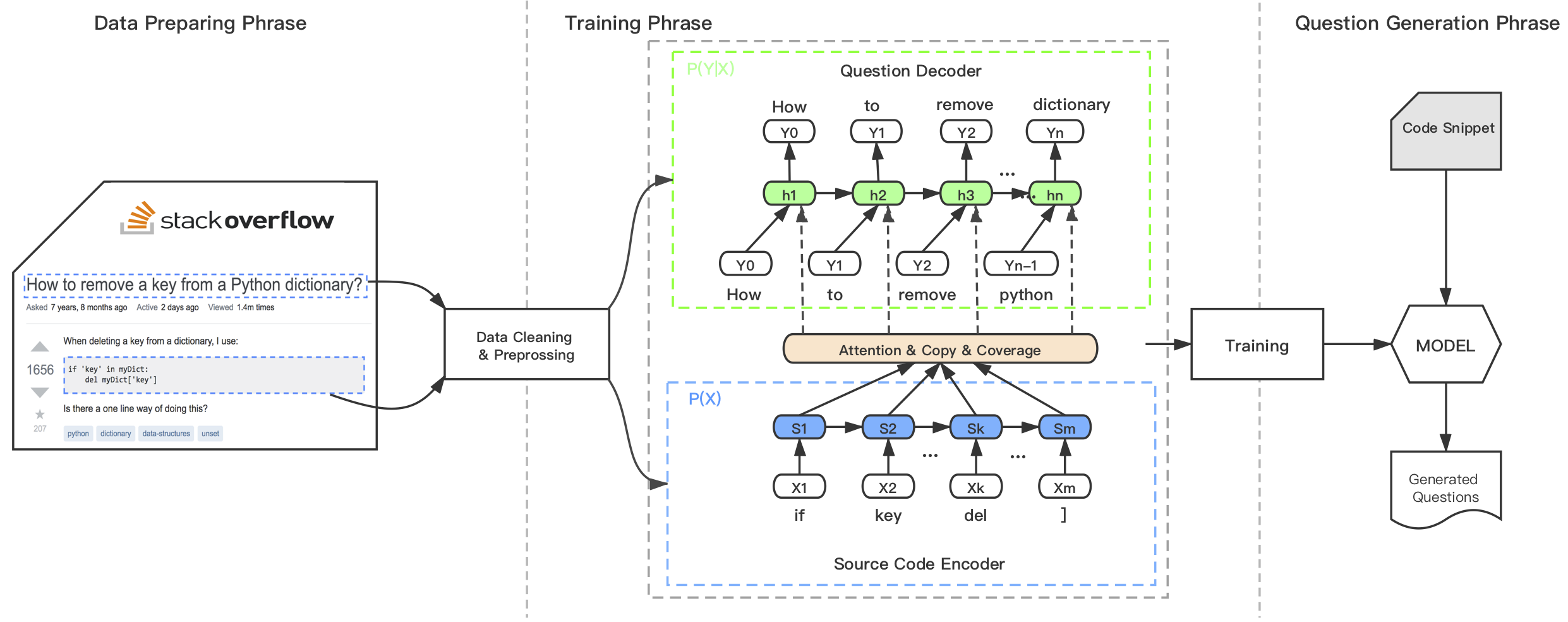}}
\vspace*{-2pt}
\caption{Workflow of Our Model}
\label{fig:workflow}
\end{figure*}

In this section, 
\rev{we firstly define the  task of question generation, then present the details of Stack Overflow question generation system.}
Fig.~\ref{fig:workflow} demonstrates the workflow used by our model. A Long Short Term Memory (LSTM) encoder-decoder architecture, is enhanced by \textit{attention} mechanism~\cite{bahdanau2014neural}, \textit{copy} mechanism~\cite{gu2016incorporating} and \textit{coverage} mechanism~\cite{tu2016modeling}. In general, our model consists of two components: \textbf{A Source-code Encoder} and \textbf{A Question Decoder}. The source code snippet is transformed by Source-code Encoder into a vector  representation, which is then read by a Question Decoder to generate the target \rev{question titles}.
Our model is a differentiable Seq2Seq model with aforementioned three mechanism,  i.e., \textit{attention} mechanism, \textit{copy} mechanism and \textit{coverage} mechanism, which can be trained in an end-to-end fashion with gradient descent.

\subsection{\rev{Question Generation Task Definition}}
\rev{
The motivation for our work is to improve the low-quality questions in Stack Overflow. 
Considering many developers may not be able to describe the problems due to their lack of knowledge and terminology, and/or they are not native english speakers, we propose a novel task in this paper - automatic generation of question titles from a code snippet, the central theme of which is helping developers to create better question titles based on their targets and code snippets. We formulate this task as a sequence-to-sequence learning problem.
}

\rev{
Given $\mathbf{C}$ is the sequence of tokens within a code snippet, our target is to generate a Question $\mathbf{Q}$, which is relevant, natural, syntactically and semantically correct. To be more specific, our main objective is to learn the underlying conditional probability distribution $P_{\theta}(\mathbf{Q}|\mathbf{C})$ parameterized by $\theta$. In other words, the goal is to train a model $\theta$ using $\langle$\textit{code snippet, \textit{question}}$\rangle$ pairs such that the probability $P_{\theta}(\mathbf{Q}|\mathbf{C})$ is maximized over the given training dataset.
More formally given a code snippet $\textbf{C}$ as a sequence of tokens $(x_1, x_2, ..., x_{M})$ of length $M$, and a question title $\textbf{Q}$ as a sequence of natural language words $(y_1, y_2, ..., y_{N})$ of length $N$. Mathematically, our task is defined as finding $\overline{y}$, such that:
\begin{equation}
    \overline{y} = argmax_{\mathbf{Q}}P_{\theta}(\mathbf{Q}|\mathbf{C})
\end{equation}
where $P_{\theta}(\mathbf{Q}|\mathbf{C})$ is defined as:
\begin{equation}
    P_{\theta}(\mathbf{Q}|\mathbf{C}) = \prod^{L}_{i=1}P_{\theta}(y_i|y_1, ..., y_{i-1};x_1, ..., x_M)
\end{equation}
$P_{\theta}(\mathbf{Q}|\mathbf{C})$ can be seen as the conditional log-likelihood of the predicted question title $\mathbf{Q}$ given the input code snippet $\mathbf{C}$.
}


\subsection{Source-code Encoder}
Source code token in the code snippet is fed sequentially into the encoder, which generates a sequence of hidden states. Our encoder is a two-layer bidirectional LSTM network,
\begin{equation}
\nonumber
\begin{split}
\overrightarrow{\mathbf{f}{w}_{t}} &= \overrightarrow{{\textnormal{LSTM}_2}} \left( x_{t}, \overrightarrow{\mathbf{h}_{t-1}}\right) \\
\overleftarrow{\mathbf{b}{w}_{t}} &= \overleftarrow{{\textnormal{LSTM}_2}} \left( x_{t}, \overleftarrow{\mathbf{h}_{t-1}}\right)
\end{split}
\end{equation}
where $x_{t}$ is the given input source code token at time step step $t$, and $\overrightarrow{\mathbf{h}_{t}}$ and $\overleftarrow{\mathbf{h}_{t}}$ are the hidden states at time step $t$ for the forward pass and backward pass respectively. The hidden states(from the forward and backward pass) of the last layer of the source-code encoder are concatenated to form a state $s$ as  $\mathbf{s} = [\overrightarrow{\mathbf{f}{w}_{t}}; \overleftarrow{\mathbf{b}{w}_t} ]$.

\subsection{Question Decoder}
Our question decoder is a singe-layer LSTM network, initialized with the state $s$ as  $\mathbf{s} = [\overrightarrow{\mathbf{f}{w}_{t}}; \overleftarrow{\mathbf{b}{w}_t} ]$. Let $qword_{t}$ be the target word at time stamp t \emph{of the ground truth \rev{question title}}. During training, at each time step $t$ the decoder takes as input the embedding vector $y_{t-1}$ of the previous word $qword_{t-1}$ and the previous state $s_{t-1}$, and concatenates them to produce the input of the LSTM network. The output of the LSTM network is regarded as the decoder hidden state $s_t$, as follows:
\begin{equation}
\mathbf{s}_t = \textnormal{LSTM}_1 \left( y_{t-1} , \mathbf{s}_{t-1}\right)
\end{equation}
The decoder produces one symbol at a time and stops when the END symbol is emitted. The only change with the decoder at testing time is that it uses output from the previous word emitted by the decoder in place of $word_{t-1}$ (since there is no access to a ground truth then).

\subsection{Incorporating Attention Mechanism}

We model the attention ~\cite{bahdanau2014neural} distribution over words in the source code snippets. We calculate the attention $(a^{t}_{i})$ over the $i^{th}$ code snippet token as :

\begin{equation}
    e^{t}_{i} = v^{t}\textnormal{tanh}\left(W_{eh}h_{i} + W_{sh}s_{t} + b_{att} \right)
\end{equation}
\begin{equation}
     a^{t}_{i} = \textnormal{softmax} \left( e^{t}_{i} \right)
\end{equation}
Here, $v^{t}$, $W_{sh}$ and $b_{att}$ are model parameters to be learned, and $h_{i}$ is the concatenation of forward and backward hidden states of source-code encoder. We use this attention $a^{t}_{i}$ to generate the context vector $c^{*}_{t}$ as the weighted sum of encoder hidden states :

\begin{equation}
\mathbf{c}^{*}_{t} = \sum_{i=1,..,|\mathbf{x}|} a^{t}_{i} \mathbf{h}_i
\end{equation}

We further use the $c^{*}_{t}$ vector to obtain a probability distribution over the words in the vocabulary as follows,

\begin{equation}
P = \textnormal{softmax} \left(\mathbf{W}_{v}[s_{t}, c^{*}_{t}] + b_{v} \right)
\end{equation}
where $W_{v}$ and $b_{v}$ are model parameters. Thus during decoding, the probability of a word is $P(qword)$. During the training process for each word at each timestamp, the loss associated with the generated \rev{question title} is :

\begin{equation}
Loss = -\frac{1}{T} \sum^{T}_{t=0}logP(qword_{t})
\end{equation}
\rev{The \textit{attention} mechanism allows the model to focus on the most relevant parts of the input sequence as needed. For example in Fig.~\ref{fig:workflow}, at time step 2, the context vector $c^{*}_{t}$ amplifies related hidden states $h_{k}$ with high scores, and drowning out unrelated hidden states with low scores. For such a case, it enables the question decoder to focus on the word ``del'' when it generates the word ``remove''. This ability to amplify the signal from the relevant part of the input sequence makes attention models produce better results than models without attention.
}

\subsection{Incorporating Copy Mechanism}
A \textit{copy} mechanism ~\cite{gu2016incorporating} is used to facilitate copying some tokens from the source code snippet to the target generated \rev{question title}. As illustrated in Fig.~\ref{fig:example}, some words such as ``setUpClass'' are naturally going to be much less frequent than other words. Thus it is highly unlikely for a decoder that is solely based on a language model to generate such a word with very rare occurrences in a corpus. In such cases, the possibly rare words in the input sequence might be required to be $copied$ from our source code snippet to the target generated \rev{question title}. We incorporate a \textit{copy} mechanism to handle such rare word problem for Stack Overflow question generation.

\begin{figure}\vspace{-0.3cm}
\centerline{\includegraphics[width=0.72\textwidth]{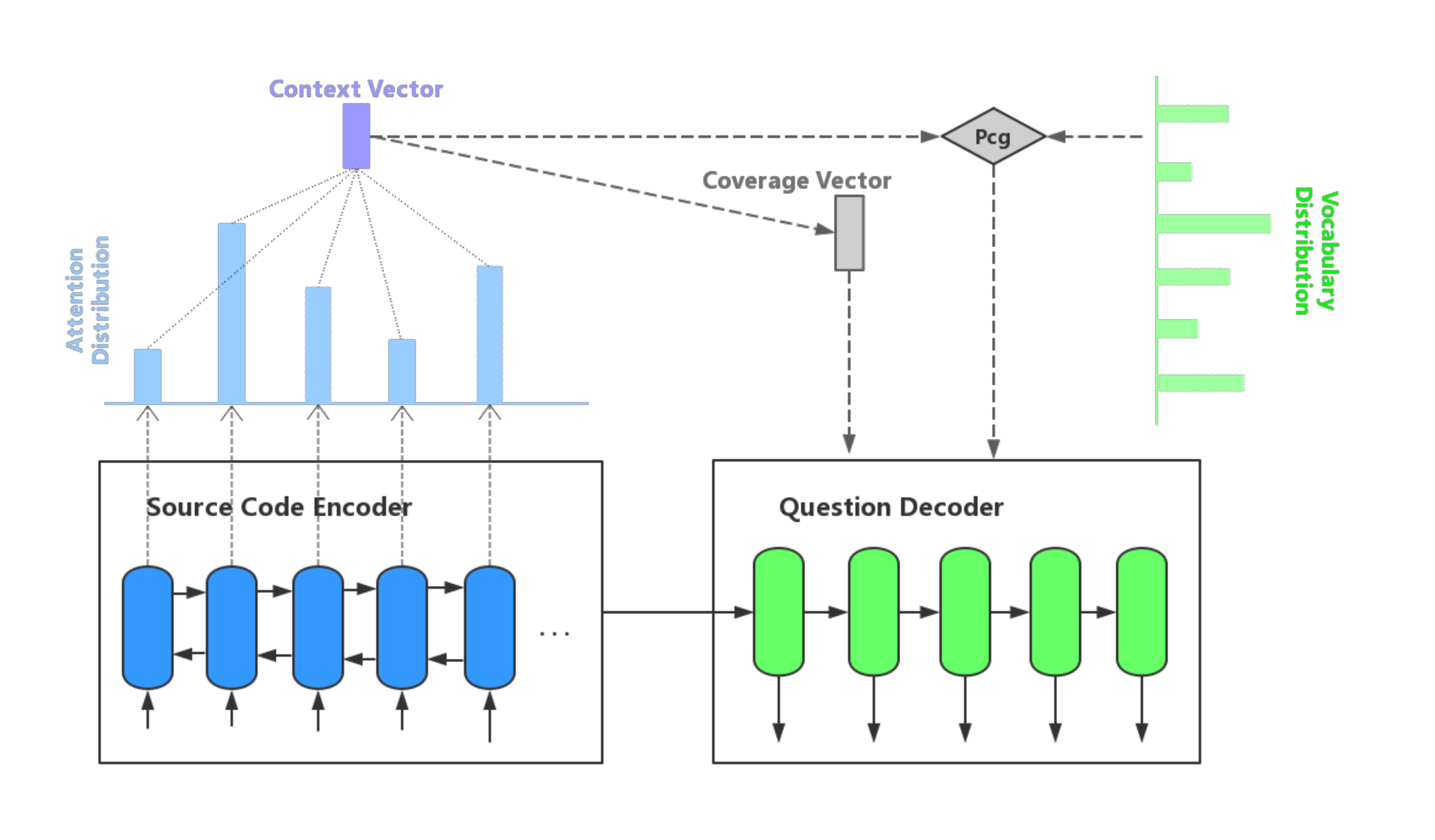}}
\vspace*{-10pt}
\caption{Attention \& Copy \& Coverage Mechanism}
\label{fig:encoder-decoder}\vspace{0.1cm}
\end{figure}

In order to learn to copy (from source) as well as to generate words from the vocabulary (using the decoder), we calculate $p_{cg} \in [0,1]$. This is the decision of a binary classifier that determines whether to generate a word from the vocabulary or to copy the word directly from the input code snippet, based on attention distribution $a^{t}_{i}$:

\begin{equation}
    p_{cg} = sigmoid(W^{T}_{eh}c^{*}_{t} + W^{T}_{sh}s_{t} + W_{x}x_{t} + b_{cg})
\end{equation}
Here $W_{eh}$, $W_{sh}$, $W_{x}$ and $b_{cg}$ are trainable model parameters. The final probability of decoding a word is specified by the mixture model :
\begin{equation}
    p^{*}(qword) = p_{cg} \sum_{i:w_{i}=qword}a^{t}_{i} + (1-p_{cg})p(qword)
\end{equation}
where $p*(qword)$ is the final distribution over the union of the vocabulary and the input sequence. As discussed earlier, Equation (10) addresses the rare words issue, since a word not in our vocabulary will have probability $p(qword)=0$. Therefore, in such cases, our model will replace the $<unk>$ token for out-of-vocabulary words with a word in the input sequence having the highest attention obtained using attention distribution $a^{t}_{i}$.
\rev{
The \textit{copy} mechanism allows the model to locate a certain segment of the input sequence and puts that segment into the output sequence. $p_{cg}$ is a soft switch to choose between generating a word from vocabulary or copying a word from the input sequence.  
For example, in Fig.~\ref{fig:example}, the rare word ``setUpClass'' in the question title is copied from the input source code snippet. For such a rare word, \textit{copy} mechanism increases the copy-mode probability and decreases the generate-mode probability, which can correctly catch the rare word and put it to the output sequence.  
}

\subsection{Incorporating a Coverage Mechanism}

Repetition is a common problem for sequence-to-sequence models and to discourage meaningless repetitions, we maintain a word coverage vector $cov$,
which is the sum of attention distributions over all previous decoder timesteps:
\begin{equation}
    cov^{t} = \sum^{t-1}_{t'=0}a^{t'}
\end{equation}
Intuitively, $cov^{t}$ is a distribution over source code snippet tokens that represents the degree of coverage that those tokens have received from the \textit{attention} mechanism so far. Note that no word is generated before timestamp 0, and hence $cov^{0}$ will be a zero vector then. The update equation (4) is now modified to be:
\begin{equation}
    e^{t}_{i} = v^{t}\textnormal{tanh}\left(W_{cv}cov_{i}^{t} + W_{eh}h_{i} + W_{sh}s_{t} + b_{att} \right)
\end{equation}
Here, $W_{cv}$ are trainable parameters that ensure the \textit{attention} mechanism's current decision is informed by a reminder of its previous decisions.
\rev{
The \textit{coverage} mechanism allows our model to solve the word repetition problem in the output sequence (see Figure~\ref{fig:ablation_example}). The \textit{coverage} mechanism ensures that the \textit{attention} mechanism's current  decision is informed by a reminder of its previous decisions (summarized in $cov^{t}$). This should make it easier for the \textit{attention} mechanism to avoid repeatedly attending to the same locations, and thus avoid generating repetitive text.
}

Following the incorporation of the copy and \textit{coverage} mechanism in our attentional sequence-to-sequence architecture, the final loss function will be:
\begin{equation}
    Loss = \frac{1}{T} \sum^{T}_{t=0}logP^{*}(qword_{t}) + \lambda L_{cov}
\end{equation}
where $\lambda$ is a reweighted hyperparameter and the coverage loss $L_{cov}$ is defined as:
\begin{equation}
    L_{cov} = \sum_{i}min(a_{i}^{t}, cov_{i}^{t})
\end{equation}
Once the model is trained, we do inference using a beam search. The beam search is parametrized by the possible paths number $k$. The inference process stops when the model generates the END token which stands for the end of the sentence.

\section{Experimental Setup}
\label{sec:eval}
In this section, we firstly describe the evaluation corpus of the task.  We then introduce the implementation details of our neural generation approach, the baselines to compare, and their experimental settings. Lastly, we explain the evaluation metrics.
 
\subsection{Pre-processing}
We experiment with our neural question generation model on the \rev{latest dump of the Stack Overflow (SO) dataset, which is publicly available\footnote{\url{https://archive.org/details/stackexchange}}. 
}
Each post comprises a short question title, a detailed question body, and one or more associated answers and multiple tags.

In this study, 
\rev{
we performed our experiment on a variety of programming languages, which include Python, Java, Javascript, C\# and SQL.
To do that, we used the \textit{Python}, \textit{Java}, \textit{Javascript}, \textit{C\#} and \textit{SQL} tag for collecting questions associated with the corresponding programming language respectively.
}
Then we removed all questions whose question scores were less than 1. This is reasonable since our goal is to generate high-quality questions to help developers.
We extracted code snippets (using $\langle$\textit{code}$\rangle$  tags) within the post's question body and corresponding post question title. We added the resulting $\langle$\textit{question}, \textit{code snippet}$\rangle$ pairs to our corpus.

\begin{table} \vspace{-0.0cm}
\rev{
\caption{\rev{Dataset Statistics}}
\begin{center}
\vspace{-0.2cm}\begin{tabular}{||l | c | c | c | c ||}
    \hline
    \ Languages & \#\emph{Code Tokens} & \#\emph{Question Tokens} & Avg.\emph{Code Length} & Avg.\emph{Question Length} \\
    \hline\hline
    \ Python & 2,367,148 & 109,329 & 84.7  & 11.2   \\
    \hline
    \ Java   & 3,371,946   & 123,994 & 103.2 & 10.8   \\
    \hline
    \ Javascript   & 2,814,729 & 121,854 & 94.1  & 10.8   \\
    \hline
    \ C\#   & 2,340,202 & 100,178 & 82.1 & 11.0   \\
    \hline
    \ SQL   & 1,483,056 & 48,668 & 84.1  & 10.1   \\
    \hline
\end{tabular}
\label{tab:datastatistics}
\end{center}
}
\vspace{-0.0cm}
\end{table}

\begin{table*} \vspace{-0.0cm}
\caption{\rev{Number of Training/Validation/Testing Samples}}
\rev{
\begin{center}
\vspace{-0.2cm}\begin{tabular}{||c|l|c|l|c||}
    \hline
    \multirow{2}{*}{Python}  
                              & \# pairs (Train) & 186,976 & \# pairs (Test-Raw) & 3,000 \\ \cline{2-5}
                              & \# pairs (Val) & 3,000 & \# pairs (Test-Clean) & 2,940 \\ \cline{2-5}
    \hline
    \multirow{2}{*}{Java}   
                              & \# pairs (Train) & 250,708 & \# pairs (Test-Raw) & 3,000 \\ \cline{2-5}
                              & \# pairs (Val) & 3,000 & \# pairs (Test-Clean) & 2,963 \\ \cline{2-5}
    \hline
    \multirow{2}{*}{Javascript}   
                              & \# pairs (Train) & 290,610 & \# pairs (Test-Raw) & 3,000 \\ \cline{2-5}
                              & \# pairs (Val) & 3,000 & \# pairs (Test-Clean) & 2,940 \\ \cline{2-5}
    \hline
    \multirow{2}{*}{C\#} 
                              & \# pairs (Train) & 178,830 & \# pairs (Test-Raw) & 3,000 \\ \cline{2-5}
                              & \# pairs (Val) & 3,000 & \# pairs (Test-Clean) & 2,974 \\ \cline{2-5}
    \hline
    \multirow{2}{*}{SQL}
                              & \# pairs (Train) & 150,002 & \# pairs (Test-Raw) & 3,000 \\ \cline{2-5}
                              & \# pairs (Val) & 3,000 & \# pairs (Test-Clean) & 2,980 \\ \cline{2-5}
    \hline
\end{tabular}
\label{tab:dataoverview}
\end{center}
}
\vspace{-0.0cm}
\end{table*}

\begin{figure}\vspace{-0.0cm}
\centerline{\includegraphics[width=0.75\textwidth]{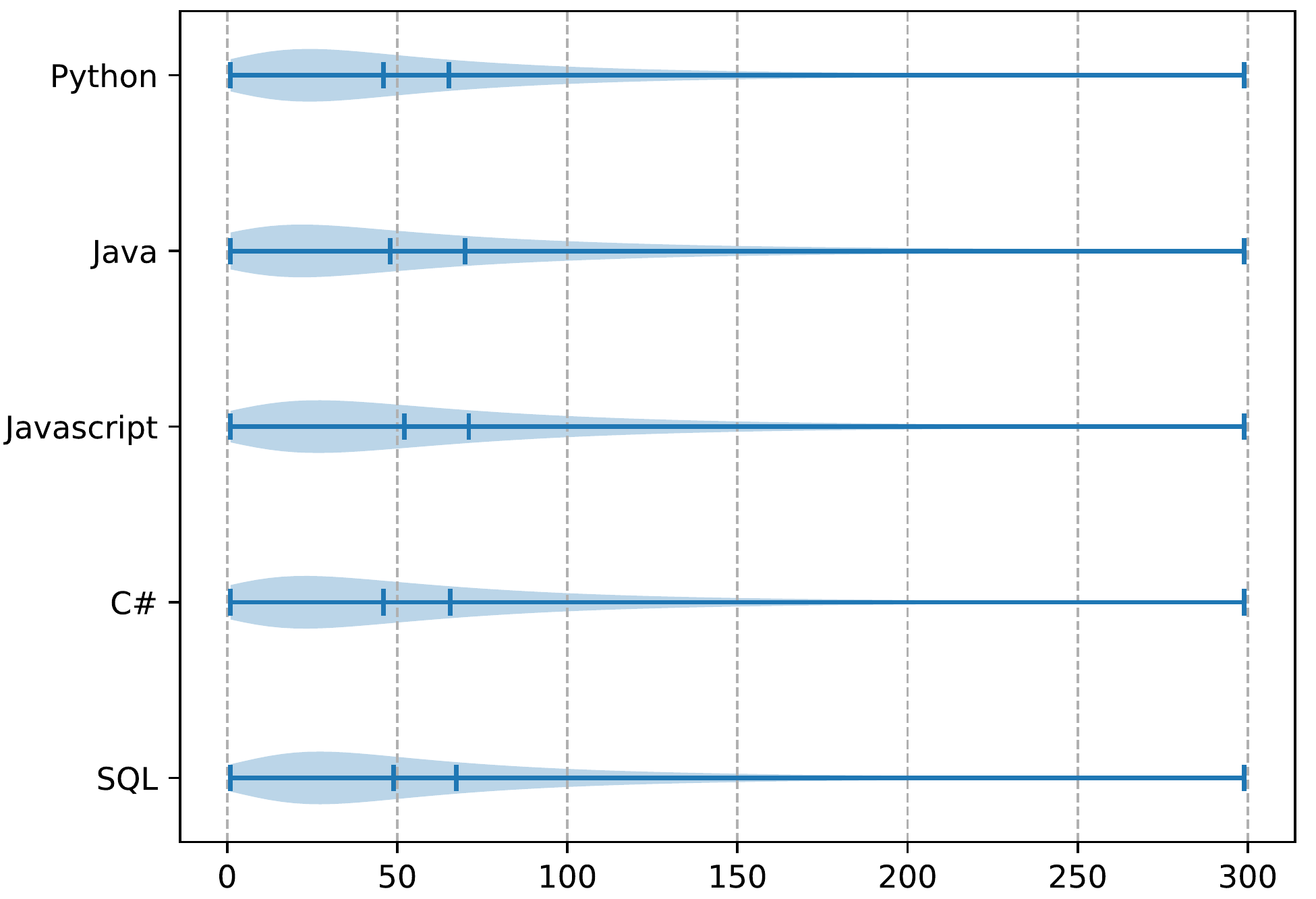}}
\vspace*{-0pt}
\caption{Volinplots of Code Distribution}
\label{fig:codedistribution}\vspace{0.1cm}
\end{figure}

\begin{figure}\vspace{-0.0cm}
\centerline{\includegraphics[width=0.75\textwidth]{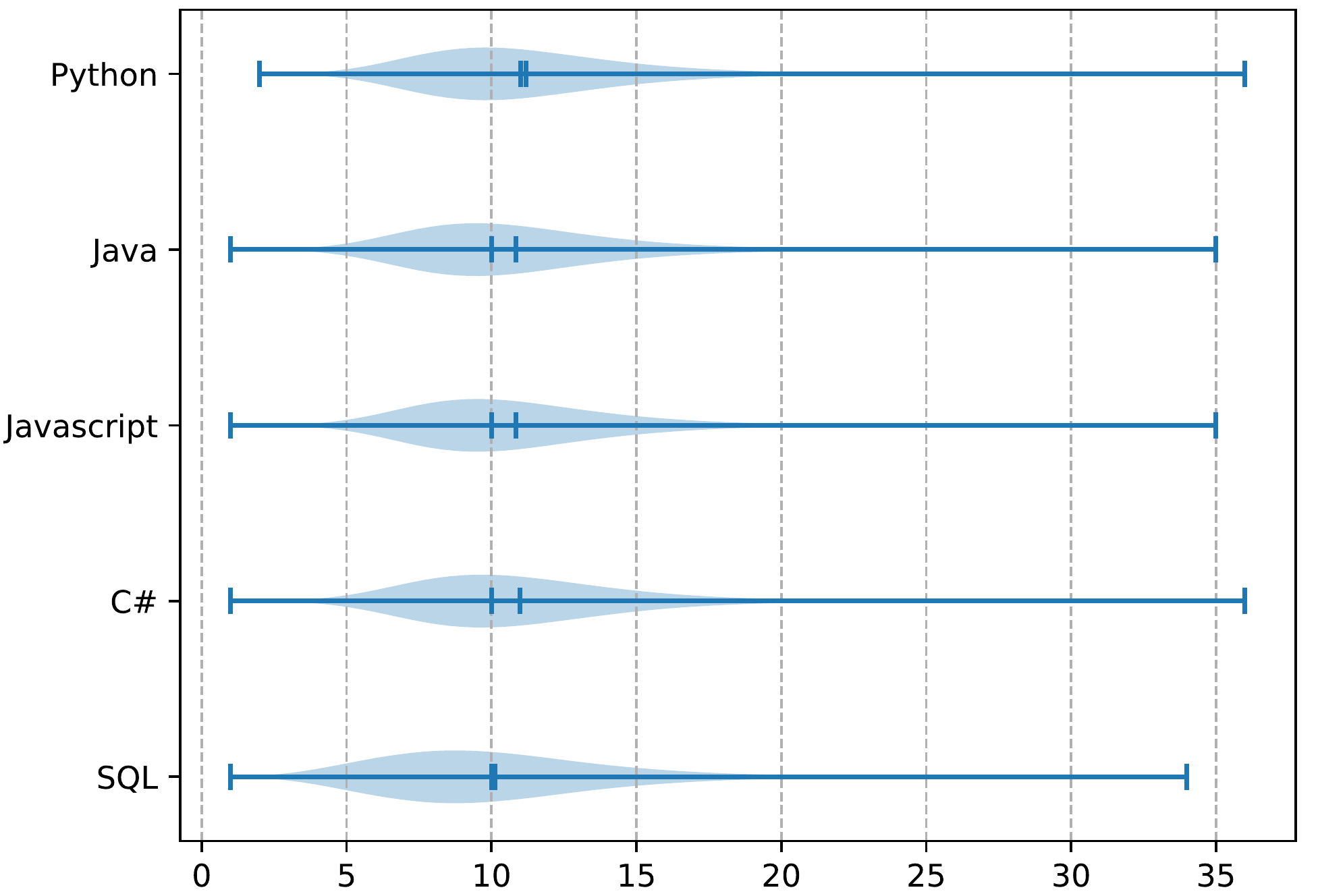}}
\vspace*{-0pt}
\caption{Volinplots of Question Distribution}
\label{fig:questiondistribution}\vspace{0.1cm}
\end{figure}

\subsubsection{Data Preprocessing}
\rev{
We tokenized the code snippet with respect to each programming language for pre-processing respectively. 
}
We adopted the NLTK toolkit~\cite{bird2004nltk} to separate tokens and symbols.
One of the challenging tasks during the tokenization was the structural complexity of the code snippet in our dataset. 
\rev{
We stripped out all comments by using the regular expression for different programming languages. After that, in order to avoid being context-specific, numbers and strings within a code snippet and replaced them with special tokens ``VAR'', ``NUMBER'' and ``STRING'' respectively. Table~\ref{tab:datastatistics}, Fig. ~\ref{fig:codedistribution} and Fig.~\ref{fig:questiondistribution} shows some data statistics on the processed dataset.
We can see that the length of Java and Javascript code snippets are much longer than the other programming languages. On average, Java and Javascript code snippets contain 103 and 94 tokens respectively, while the code snippets of the other three programming languages are just around 84 tokens long. On the other hand, the question titles of all the programming languages are approximately at the same level, the overall average of the question titles are 11 tokens long.
}

\subsubsection{Data Filtering}
Users can post different types of questions in SO, such as ``how to X'' and ``What/Why is Y''. In our preliminary study, we targeted questions which include interrogative keywords such as ``how'', ``what'', ``why'', ``which'', ``when''.
For the above collection of question-code pairs, only the pairs where the aforementioned keywords appear in the question title were kept.
\rev{
After that, we removed pairs where the code snippets are too long or too short. 
Based on the interquartile range (IQR) of the violin plots in Fig.~\ref{fig:codedistribution} and Fig.~\ref{fig:questiondistribution}, we only preserved pairs where the token range from 16 tokens to 128 tokens for code snippet and the token range from 4 tokens to 16 tokens for question titles. 
At this stage, we collected more than 1M $\langle$\textit{question}, \textit{code snippet}$\rangle$ pairs in total for Python, Java, Javascript, C\# and SQL programming languages. We randomly sampled 3,000 pairs for validation and 3,000 pairs for testing respectively, and kept the rest for training. The details of the training, validation and testing samples for each programming language are summarized in Table~\ref{tab:dataoverview}.
}

\subsubsection{\rev{Clone Detection}}
\rev{
Considering that there may be duplicate and/or very similar $\langle$\textit{code snippet}, \textit{question}$\rangle$ pairs between the training set and testing set, this may mislead the evaluation results. We further conducted a primitive clone detection analysis to remove the noisy examples from our testing data set. A lot of methods have been proposed for clone detection in recent years (e.g., ~\cite{wang2020detecting, gao2020checking, white2016deep, buch2019learning, gao2019smartembed}). We followed the approach proposed by \cite{gao2019smartembed} for clone detection. For each code snippet, we compose a numerical vector by summing up the word embedding vectors for all the relevant tokens within the code snippet. Then the similarity between two code snippets $C_1$ and $C_2$ can be calculated as follows:
\begin{equation}
Distance(C_1, C_2)= Euclidean(e_1, e_2)
\label{eq2}
\end{equation}
\begin{equation}
Similarity(C_1, C_2)= 1 - Distance(C_1, C_2) \label{eq3}
\end{equation}
where $e_1$ and $e_2$ are the corresponding code embedding vectors of $C_1$ and $C_2$. Each code snippet $C_i$ in the testing set is queried against all the code snippets in the training set, the maximum similarity score $s_i$ associated with the $C_i$ is retrieved. The results of $s_i$ with respect to each programming language are summarized in Table~\ref{tab:clonedetection}. 
If the similarity score $s_i$ is over a threshold $\delta$ ($\delta$ is set to 0.8 in this study), then the code snippet $C_i$ is viewed as a code clone and will be deleted from our testing set. From the table we can see that the number of clone code snippets is very small, while most code snippets get relatively low similarity scores. After removing all the examples with similarity scores above 0.8 from the testing set, we reconstructed a clean testing set for each programming language, the final results are summarized in Table~\ref{tab:dataoverview}. The clean testing set is used for the final evaluation of this study. 
}

\begin{table} \vspace{-0.0cm}
\rev{
\caption{\rev{Clone Detection Analysis}}
\begin{center}
\vspace{-0.2cm}\begin{tabular}{|| c | c | c | c | c | c ||}
    \hline
    \ Similarity & Python & Java & Javascript & C\# & SQL \\
    \hline\hline
    $s_i \in [0.0, 0.2)$   & 2,153 & 2,241 & 1,939  & 2,328 & 2,359  \\
    \hline
    $s_i \in [0.2, 0.4)$   & 512   & 473 & 651 & 422 & 413  \\
    \hline
    $s_i \in [0.4, 0.6)$   & 195   & 187 & 272 & 182 & 159  \\
    \hline
    $s_i \in [0.6, 0.8)$   & 80    & 62 & 78 & 42 & 49  \\
    \hline
    $s_i \in [0.8, 1.0]$   & 60    & 37 & 60 & 26 & 20  \\
    \hline
\end{tabular}
\label{tab:clonedetection}
\end{center}
}
\vspace{-0.0cm}
\end{table}

\subsection{Implementation Details}
We implemented our system in Python using Tensorflow framework.
We added special START and END tokens for each sequence in our training set.
The vocabulary size for the Java and Python dataset were set to 50,000 and 80,000 respectively.
We use a two-layer bidirectional LSTM for the encoder and a single-layer LSTM for the decoder. We set the number of LSTM hidden states to 256 in both encoder and decoder. We choose the word embeddings of 300 dimensions.
Optimization is performed using stochastic gradient descent (SGD) with a learning rate of 0.01. We fix the batch size for updating to be 32. During decoding, we perform beam search with beam size of 10.
We train the model for 30 epochs. Our hyper-parameters were tuned on the validation set, the evaluation results were reported on the test set.
We discuss the details of the parameter tuning in Section~\ref{sec:results}.

\subsection{Baselines}
To demonstrate the effectiveness of our proposed approach, we compared it with several competitive baseline approaches. We adapted these approaches slightly for our problems, i.e., generating \rev{question titles} from a given code snippet. We briefly introduced these approaches and the experimental settings as below. For each method mentioned below, the involved parameters were carefully tuned, and the parameters with the best performance were used to report the final comparison results.
\begin{enumerate}
    \item \textbf{IR} stands for the information retrieval baseline. For a given code snippet $c_i$, it retrieves the \textit{question titles} associated with the code $c_j$ that is closest to the input code $c_i$ from the training set. We use TF-IDF~\cite{robertson1994some} metric to calculate the distance between two code snippets, and build a nearest neighbor model to retrieve the most similar instance from the traning set.
    \item \textbf{MOSES} ~\cite{koehn2007moses} is a widely used phrase-based statistical machine translation system. Here, we treat a tokenzied code snippet as the source language text, and the corresponding question title as the target language text. We run the translation from code snippets to \rev{question titles}. We train a 3-gram language model on target side texts using KenLM~\cite{heafield2011kenlm}, and perform turning with MERT on dev set.
    \item \textbf{NMT} Jiang et al.~\cite{jiang2017automatically} proposed an sequence-to-sequence approach to generate commit message from code, we refer to it as NMT in our study. We choose NMT as one comparing approach since its promising performance in commit generation. NMT model take source code as inputs and associated \rev{question title} as outputs. Hyperparameters are tuned with validation set.
    \rev{
    \item \textbf{CODE-NN} Iyer et al.~\cite{iyer2016summarizing} proposed an attention-based Long Short Term Memory (LSTM) neural network, named CODE-NN, to generate descriptive summaries for C\# code snippets and SQL queries. In order to use CODE-NN, the C\# code fragments and SQL statements first need to be parsed by the modified version of parser. Considering code snippets in SO are usually incomplete and not parsable, and it is non-trivial to design specific parser to parse code snippets of various programming languages, we tried our best to apply our approach to the CODE-NN dataset, which include 60k+ C\# (title, query) pairs and 30k+ SQL (title, query) pairs respectively. 
    }
\end{enumerate}

\subsection{Evaluation Metrics}

We evaluate our task with automatic evaluation, and also perform human evaluation via a user study.
\begin{enumerate}
    \item \textbf{Automatic Evaluation} To evaluate different models, We adopt BLEU-1, BLEU-2, BLEU-3, BLEU-4~\cite{papineni2002bleu}, ROUGE-1, ROUGE-2 and ROUGE-L~\cite{lin2004rouge} scores.
    BLEU is a precision-oriented measure commonly used in translation tasks, which measures the average $n$-gram precision on a set of reference sentences, with a penalty for overly short sentences. BLEU-$n$ is the BLEU score that uses up to $n$-grams for counting co-occurrences.
    ROUGE is a recall-oriented measure widely used in summarization tasks, which used to evaluate $n$-grams recall of the summaries with gold-standard sentences as references.
    ROUGE-1 and ROUGE-2 measures the uigram and bigrams between the system and reference summaries.
    ROUGE-L is a longest common subsequence measure metric, it does not require consecutive matches but in-sequence matches that reflect sentence level word order. 
    \rev{
    We conducted a large scale automatic evaluation over various kinds of programming languages, i.e., Python, Java, Javascript, C\# and SQL. 
    In our work, we regard the generated \textit{question titles} as candidates, and the original human written \rev{question titles} as gold-standard references.
    }
    \item \textbf{Human Evaluation}Since automatic evaluation of generated text does not always agree with the actual human-perceived quality and usefulness of the results, we also perform human evaluation studies to measure how humans perceive the generated questions. To do this, we consider two modalities in our user study : \textit{Naturalness} and \textit{Relevance}. \textit{Naturalness} measures the grammatical correctness and fluency of the \rev{question title} generated. \textit{Relevance} measures how relevant the \textit{question title} is to the code snippet, and indicates the factual divergence of the code snippet to the reference question titles. 
    \rev{
    We randomly sampled 50 $\langle$\textit{code snippet}, \textit{question}$\rangle$ pairs from Python and Java test results respectively, 
    }
    for each code snippet, we provided 5 associated \textit{question titles}: one was generated by human (the ground truth \textit{question title}), while the others were generated by baseline methods and our approach. Then we invited 5 evaluators, including 4 Ph.D students and 1 Masters student, all of whom are not co-authors, majoring in Computer Science and have industrial experience with Python as well as Java programming (ranging from 1-3 years). 
    \rev{All of the five evaluators have at least one year of studying/working-experience in English speaking countries. 
    }
    Each participant was asked to manually rate generated \rev{question titles} on a scale between 1 and 5 (5 for the best results) across the above modalities. The volunteers were blinded as to which \rev{question title} was generated by our approach.
    \rev{
    \item \textbf{Practical Manual Evaluation} Following the human evaluation, we also performed a practical manual evaluation to further analyze whether our approach can generate better question titles for \textit{low-quality} questions in Stack Overflow. To do this, we randomly sampled 50 \textit{low-quality} $\langle$\textit{code snippet}, \textit{question}$\rangle$ pairs from our Python and Java datasets before the data preprocessing. It is worth mentioning that different from human evaluation, these sampled posts were not included in our training and/or testing set, because all the questions with score less than 1 were removed before training processing. For each code snippet, we applied our approach to generate a question title for manual annotation. We conducted pairwise comparison between two question titles (one was generated by humans, one was generated by our tool) for the same code snippet. For each pairwise comparison, we asked the same 5 evaluators to decide which one is better or non-distinguishable in terms of the following three metrics: \textit{Clearness}, \textit{Fitness}, \textit{Willingness to Respond}.
    \textit{Clearness} measures whether a question title is expressed in a clear way. Unclear questions are ambiguous, vague, and/or incomplete.
    \textit{Fitness} measures whether a question title is reasonable in logic with the provided code snippet, and whether it is questioning on the key information. Unfit question titles are either irrelevant to the code snippet or universal questions. 
    \textit{Willingness to Respond} measures whether a user is willing to respond to a specific question. This metric is used to justify how likely the generated questions can elicit further interactions. If people are willing to respond, the interactions can go further. Each metric is evaluated independently on each pairwise comparison. Also the two question titles were randomly shuffled and the participants do not know which question is generated by our approach. 
    }
\end{enumerate}

\section{Results and Analysis}
\label{sec:results}
To gain a deeper understanding of the performance of our approach, we conduct analysis on our evaluation results in this section. For quantitative analysis, firstly we study the experimental results of automatic evaluation, then we examine the outcome of human evaluation. Specifically, we mainly focus on the following research questions:
\begin{itemize}
    \item \textit{RQ-1:} How effective is our approach under automatic evaluation?
    \rev{
    \item \textit{RQ-2:} How effective is our approach compared with the CODE-NN model?
    }
    \item \textit{RQ-3:} How effective is our approach under human evaluation?
    \rev{
    \item \textit{RQ-4:} How effective is our approach for improving low-quality questions? 
    }
    \item \textit{RQ-5:} \rev{How effective is our use of \textit{attention} mechanism, \textit{copy} mechanism and \textit{coverage} mechanism under automatic evaluation?}
    \item \textit{RQ-6:} How effective is our approach under different parameter settings?
    \rev{
    \item \textit{RQ-7:} How efficient is our approach in practical usage?
    }
\end{itemize}

\subsection{\textbf{RQ-1: How effective is our approach under automatic evaluation?}}
\subsubsection{Automatic Evaluation Results}
\rev{
The automatic evaluation results of our proposed model and aforementioned baselines are summarized in Table~\ref{tab:automaticevaluation_python}, ~\ref{tab:automaticevaluation_java}, ~\ref{tab:automaticevaluation_javascript}, ~\ref{tab:automaticevaluation_csharp}, ~\ref{tab:automaticevaluation_sql} for Python, Java, Javascript, C\#, and SQL respectively. The best performing system for each column is highlighted in boldface. As can be seen, \textbf{our model outperforms all the other methods considerably} in terms of BLEU score and ROUGE score.
}
BLEU score measures precision of the system. To be more specific, it measures how many words (and/or n-grams) in the machine generated \rev{question titles} appear in the ground truth \rev{question titles}. For ROUGE scores, it measures the recall of the system i.e. how many words(and/or n-grams) in the ground-truth \rev{question titles} appear in the machine generated \rev{questions titles}. From the table, we can observe the following points:

\begin{table*}
\centering
\caption{Automatic evaluation(Python dataset)}
\label{tab:automaticevaluation_python}
\vspace*{-5pt}
\resizebox{0.95\textwidth}{!}{
\rev{
    \begin{tabular}{||l|cccc|ccc||}
      \hline
      Model & BLEU-1  & BLEU-2 & BLEU-3 & BLEU-4 & ROUGE-1 & ROUGE-2 & ROUGE-L \\
      \hline
      IR\textsubscript{TFIDF} & $20.2\pm1.1\%$
                              & $17.7\pm0.4\%$
                              & $18.4\pm0.3\%$
                              & $18.0\pm0.2\%$
                              & $24.4\pm1.4\%$
                              & $6.9\pm0.6\%$
                              & $21.8\pm1.2\%$ \\
      Moses & $20.4\pm1.4\%$
            & $18.1\pm0.8\%$
            & $17.8\pm0.7\%$
            & $17.4\pm0.6\%$
            & $26.9\pm1.3\%$
            & $6.2\pm0.5\%$
            & $20.4\pm1.1\%$ \\
      NMT  & $28.9\pm1.7\%$
               & $21.9\pm0.7\%$
               & $21.3\pm0.3\%$
               & $20.3\pm0.2\%$
               & $34.1\pm2.2\%$
               & $10.6\pm1.1\%$
               & $31.2\pm1.9\%$ \\
      \hline
      \textbf{Ours}  & $\mathbf{35.8\pm2.0\%}$
                     & $\mathbf{30.1\pm0.9\%}$
                     & $\mathbf{26.8\pm0.4\%}$
                     & $\mathbf{24.2\pm0.3\%}$
                     & $\mathbf{39.9\pm2.5\%}$
                     & $\mathbf{12.6\pm2.5\%}$
                     & $\mathbf{36.7\pm2.4\%}$ \\
      \hline
\end{tabular}
}
}
\end{table*}

\begin{table*}
\centering
\caption{Automatic evaluation(Java dataset)}
\label{tab:automaticevaluation_java}
\vspace*{-6pt}
\resizebox{0.95\textwidth}{!}{
\rev{
    \begin{tabular}{||l|cccc|ccc||}
      \hline
      Model & BLEU-1  & BLEU-2 & BLEU-3 & BLEU-4 & ROUGE-1 & ROUGE-2 & ROUGE-L \\
      \hline
      IR\textsubscript{TFIDF} & $18.1\pm1.1\%$
                              & $17.2\pm0.5\%$
                              & $18.0\pm0.4\%$
                              & $17.6\pm0.3\%$
                              & $22.2\pm1.3\%$
                              & $6.2\pm0.7\%$
                              & $19.9\pm1.2\%$ \\
      Moses & $18.5\pm1.0\%$
            & $17.3\pm0.6\%$
            & $17.1\pm0.5\%$
            & $16.7\pm0.4\%$
            & $25.2\pm1.5\%$
            & $5.3\pm0.4\%$
            & $20.6\pm1.2\%$ \\
      NMT  & $25.0\pm1.6\%$
               & $20.7\pm0.7\%$
               & $20.9\pm0.3\%$
               & $20.2\pm0.2\%$
               & $30.0\pm2.0\%$
               & $9.6\pm1.1\%$
               & $27.3\pm1.8\%$ \\
      \hline
      \textbf{Ours}  & $\mathbf{31.8\pm1.8\%}$
                     & $\mathbf{27.5\pm0.7\%}$
                     & $\mathbf{25.2\pm0.3\%}$
                     & $\mathbf{23.3\pm0.2\%}$
                     & $\mathbf{35.4\pm2.2\%}$
                     & $\mathbf{10.0\pm1.8\%}$
                     & $\mathbf{32.6\pm2.1\%}$ \\
      \hline
\end{tabular}
}
}
\end{table*}

\begin{table*}
\centering
\caption{Automatic evaluation(Javascript dataset)}
\label{tab:automaticevaluation_javascript}
\vspace*{-6pt}
\resizebox{0.95\textwidth}{!}{
\rev{
    \begin{tabular}{||l|cccc|ccc||}
      \hline
      Model & BLEU-1  & BLEU-2 & BLEU-3 & BLEU-4 & ROUGE-1 & ROUGE-2 & ROUGE-L \\
      \hline
      IR\textsubscript{TFIDF} & $18.7\pm1.1\%$
                              & $17.6\pm0.4\%$
                              & $18.3\pm0.3\%$
                              & $17.9\pm0.2\%$
                              & $22.6\pm1.3\%$
                              & $6.2\pm0.6\%$
                              & $20.2\pm1.1\%$ \\
      Moses & $18.9\pm1.2\%$
            & $18.8\pm0.7\%$
            & $18.7\pm0.7\%$
            & $18.3\pm0.6\%$
            & $25.7\pm1.2\%$
            & $5.8\pm0.4\%$
            & $20.1\pm1.0\%$ \\
      NMT  & $28.1\pm1.6\%$
               & $22.0\pm0.6\%$
               & $21.5\pm0.3\%$
               & $20.5\pm0.2\%$
               & $32.8\pm1.9\%$
               & $10.3\pm1.0\%$
               & $30.4\pm1.7\%$ \\
      \hline
      \textbf{Ours}  & $\mathbf{33.2\pm1.9\%}$
                     & $\mathbf{26.4\pm0.8\%}$
                     & $\mathbf{24.1\pm0.4\%}$
                     & $\mathbf{22.1\pm0.3\%}$
                     & $\mathbf{37.3\pm2.2\%}$
                     & $\mathbf{11.7\pm1.8\%}$
                     & $\mathbf{34.7\pm2.1\%}$ \\
      \hline
\end{tabular}
}
}
\end{table*}

\begin{table*}
\centering
\caption{Automatic evaluation(C\# dataset)}
\label{tab:automaticevaluation_csharp}
\vspace*{-6pt}
\resizebox{0.95\textwidth}{!}{
\rev{
    \begin{tabular}{||l|cccc|ccc||}
      \hline
      Model & BLEU-1  & BLEU-2 & BLEU-3 & BLEU-4 & ROUGE-1 & ROUGE-2 & ROUGE-L \\
      \hline
      IR\textsubscript{TFIDF} & $18.0\pm1.0\%$
                              & $17.1\pm0.4\%$
                              & $17.9\pm0.3\%$
                              & $17.6\pm0.2\%$
                              & $21.9\pm1.3\%$
                              & $6.3\pm0.6\%$
                              & $19.9\pm1.1\%$ \\
      Moses & $18.5\pm1.0\%$
            & $16.8\pm0.7\%$
            & $16.6\pm0.6\%$
            & $16.3\pm0.6\%$
            & $25.4\pm1.2\%$
            & $6.0\pm0.4\%$
            & $20.0\pm1.0\%$ \\
      NMT  & $24.4\pm1.7\%$
               & $19.3\pm0.7\%$
               & $19.8\pm0.2\%$
               & $19.3\pm0.2\%$
               & $29.4\pm1.6\%$
               & $9.7\pm0.8\%$
               & $27.1\pm1.4\%$ \\
      \hline
      \textbf{Ours}  & $\mathbf{30.9\pm1.8\%}$
                     & $\mathbf{27.7\pm0.7\%}$
                     & $\mathbf{25.3\pm0.3\%}$
                     & $\mathbf{23.4\pm0.2\%}$
                     & $\mathbf{34.8\pm2.3\%}$
                     & $\mathbf{10.2\pm1.9\%}$
                     & $\mathbf{31.8\pm2.2\%}$ \\
      \hline
\end{tabular}
}
}
\end{table*}

\begin{table*}
\centering
\caption{Automatic evaluation(SQL dataset)}
\label{tab:automaticevaluation_sql}
\vspace*{-6pt}
\resizebox{0.95\textwidth}{!}{
\rev{
    \begin{tabular}{||l|cccc|ccc||}
      \hline
      Model & BLEU-1  & BLEU-2 & BLEU-3 & BLEU-4 & ROUGE-1 & ROUGE-2 & ROUGE-L \\
      \hline
      IR\textsubscript{TFIDF} & $15.6\pm1.0\%$
                              & $17.6\pm0.4\%$
                              & $18.4\pm0.3\%$
                              & $17.9\pm0.3\%$
                              & $19.3\pm1.2\%$
                              & $3.7\pm0.6\%$
                              & $16.4\pm1.0\%$ \\
      Moses & $17.3\pm0.9\%$
            & $16.6\pm0.7\%$
            & $16.5\pm0.6\%$
            & $16.2\pm0.6\%$
            & $21.4\pm1.1\%$
            & $3.4\pm0.3\%$
            & $15.0\pm0.8\%$ \\
      NMT  & $22.0\pm1.3\%$
               & $20.4\pm0.5\%$
               & $20.7\pm0.4\%$
               & $19.9\pm0.2\%$
               & $26.6\pm1.7\%$
               & $7.4\pm1.0\%$
               & $22.9\pm1.5\%$ \\
      \hline
      \textbf{Ours}  & $\mathbf{26.8\pm1.6\%}$
                     & $\mathbf{23.8\pm0.6\%}$
                     & $\mathbf{22.6\pm0.3\%}$
                     & $\mathbf{21.2\pm0.2\%}$
                     & $\mathbf{30.5\pm2.0\%}$
                     & $\mathbf{8.4\pm1.3\%}$
                     & $\mathbf{26.3\pm1.9\%}$ \\
      \hline
\end{tabular}
}
}
\end{table*}

\begin{enumerate}
    \item \rev{
    In general, encoder-decoder architecture baselines, i.e., NMT and our proposed methods, outperform both the IR based approach and the statistical machine translation approach (e.g., Moses) by a large margin.
    For IR based approach, it retrieves questions from existing database according to similarity score, which relies heavily on whether similar code snippets can be found and how similar the code snippets are. As a result, it is unable to consider the context of the code snippet, which is reflecting that memorizing the training set is not enough for this task.
    For the phrase-based statistical approaches which use separately engineered subcomponents, the encoder-decoder model uses the vector representation for words and internal states, semantic and structural information can be learned from these vectors by taking global context into consideration.
    }
    \item Regarding the BLEU score, 
    \rev{our approach is significantly better than the other methods (e.g., traditional IR method, phrase-based statistical method, and NMT methods) and achieves understandable results~\cite{seljan2012bleu}. 
    For example, it improves over NMT methods on BLEU-4 by \textbf{19.2\%}\footnote{The improvement ratio is defined within \url{https://www.d.umn.edu/~gshute/arch/improvements.xhtml}} on Python dataset and \textbf{15.3\%} on Java dataset.
    We attribute this to the following reasons: firstly, our approach is based on a sequence-to-sequence architecture and hence it is superior to the statistical baselines\cite{koehn2007moses}. 
    Secondly, \rev{compared with NMT baseline which is solely based on the sequence-to-sequence approach, besides using the encoder-decoder architecture, our approach} also incorporates an \textit{attention} mechanism to perform better content selection, a \textit{copy} mechanism to manage the rare-words problem in source code snippet,
    as well as a \textit{coverage} mechanism to eliminate meaningless repetitions,
    which makes it superior to the NMT baselines.
    According to ~\cite{seljan2012bleu}, the bleu-1 score above 0.30 generally reflect understandable results and above 0.50 reflect good and fluent translations, the bleu score of our approach can be considered as acceptable, but there is still a large gap compared with ground truth question titles.
    }
    \item Regarding the ROUGE score, the advantage of our proposed model is also clear.
    The potential explanation is that baseline methods, such as Moses, NMT, even with a much larger vocabulary, still has a large number of out of vocabulary words. Our model, augmented with the \textit{copy} mechanism to handle the rare-words problem, beats these baselines by a large margin. This further justifies that the \textit{copy} mechanism generally helps when dealing with the question generation tasks. It also signals that out of vocabulary tokens within code snippet convey much valuable information when generating \rev{question titles}.
    \rev{
    \item The proposed approach performs best on the Python dataset and worst on the SQL dataset. This is in part because, compared with Python code snippet, SQL code snippets only contain a set of keywords and functions, and thus generating question titles for SQL code snippet is more challenging for solely  relying on the compositional structures in the input.
    }
\end{enumerate}

\begin{figure*}
\centerline{\includegraphics[width=0.95\textwidth]{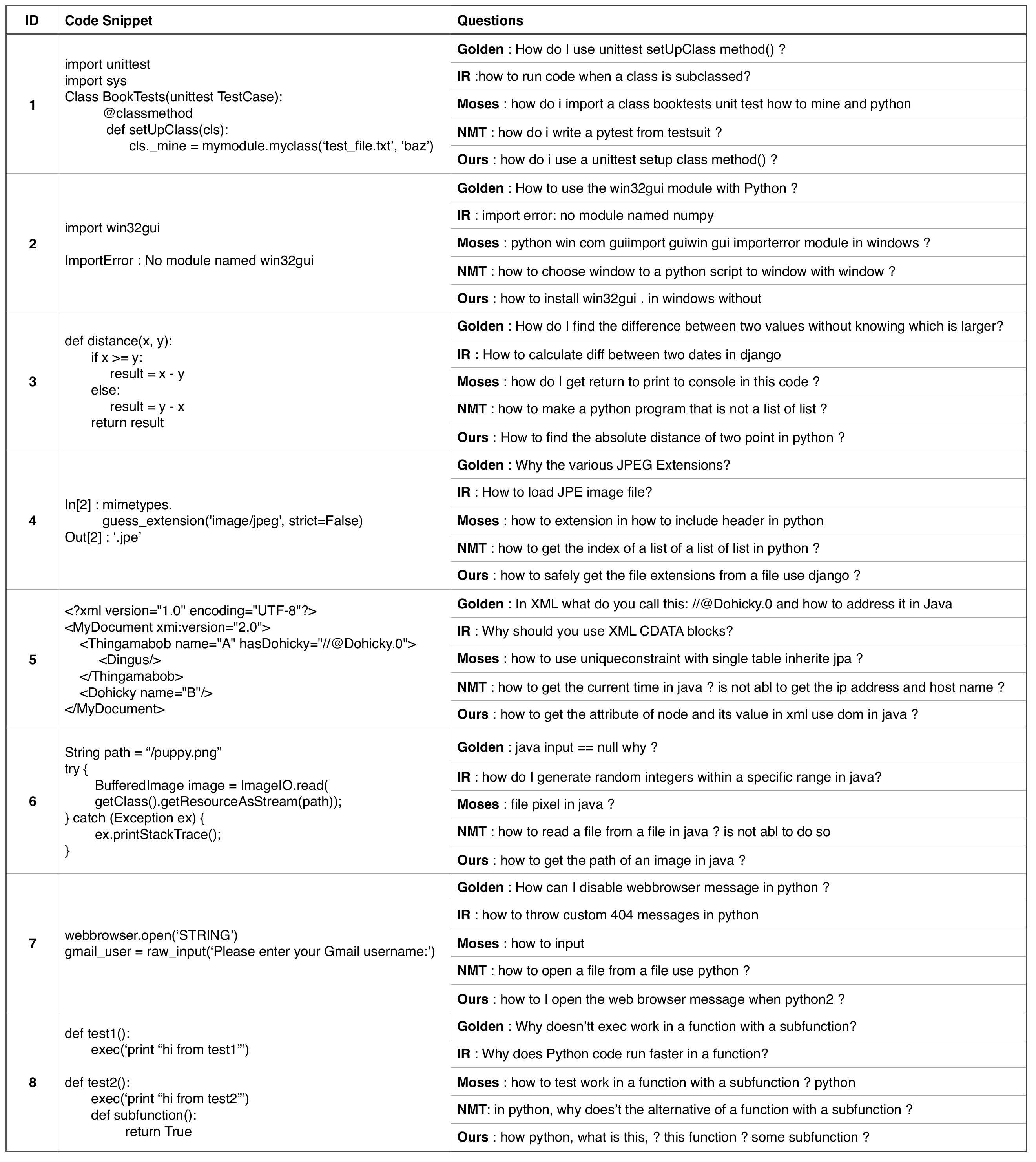}}
\caption{Examples of output generated by each model}
\label{fig:generatedexamples}
\end{figure*}

\subsubsection{\rev{Examples of the Automatic Evaluation}}

\rev{
We examine several sample outputs by hand to perform a further qualitative analysis.
Fig.~\ref{fig:generatedexamples} shows some examples of the \rev{question titles} generated by human (Golden questions), the baselines (e.g., IR, Moses and NMT) and our approach for the given code snippets in the test set. We have the following interesting observations:
\begin{enumerate}
    \item We see a large gap between our approach and other baselines. \textbf{Our approach generates syntactically and semantically correct and relevant \rev{question titles} in most cases}, while the outputs of every other model are less meaningful and/or more irrelevant. This is consistent with our previous automatic evaluation results.
    For the IR method, often the \rev{question titles} are unable to connect to the code snippet. For example in the third sample, the ground truth question is about ``find difference between two values'', while the IR methods retrieved the question of ``how to calculate the diff between two dates in django''.  
    The statistical machine translation model, such as Moses, is unable to generate a syntactically correct \rev{question title}. For example, in the sixth and seventh sample, the \rev{question titles} generated by Moses are incomplete and meaningless. 
    For the NMT method, although it can generate the \rev{question titles} in the right format in some cases, it still fails to replicate the critical tokens (e.g., example1) because of the difficulty brought by the unseen words in the code snippet.
    \item \textbf{Our approach handles out of vocabulary words well}, and it can generate acceptable \rev{question titles} for a code snippet with rare words. In contrast, the baseline methods often fail in such cases. For example, in the first sample, in which the focus should be put on ``setUpClass'' method in the code snippet, Our model successfully captures this rare phrase, while other baselines return non-relevant descriptions. It is quite interesting that our model automatically learns to select informative tokens in the code snippet, which shows the extractive ability of our model. At the same time, our approach often generates words to ``connect'' those critical tokens, showing its aspect of abstractive ability.
    \item  \textbf{A large number of the \rev{question titles} generated by our model produce meaningful output for simple code snippets}. 
    Note that in some cases, the generated \rev{question titles} are not exactly inline with the standard ones, yet still make sense by looking at the meaning of the code snippet. For example, in the second case, the ground truth \rev{question title} is ``How to use win32gui module with Python'', our system generates a \rev{question title} about ``how to install win32gui''. This is reasonable given the source code contains ``ImportError'' while ``import win32gui''.
    In the third case, our approach generates a \rev{question title} of ``how to find the absolute distance of two point in python'', this is because the code snippet defines a function that returns the distance of two points.
    For such cases, it is reasonable to generate different \rev{question titles} that look at the code snippet from different aspects. Our \rev{question titles} can also be viewed as  correct and meaningful by looking at the meanings of the code snippet.
    \item Sometimes, \textbf{our approach can generate \textit{question titles} that are more clear and informative than the ground truth \rev{question titles}}, such as samples 4-6. For example, in the fourth sample, the ground truth \rev{question title} is ``why the various JPEG extensions?'' which is uninformative and unclear to the potential helpers, after using our tool the \rev{question title} can be rephrased as ``how to safely get the file extensions from a file'' which is more attractive and informative than the original ones.
    \item However, \textbf{outputs from our system are not always ``correct''}.  For example, in the last second sample, the ground truth \rev{question title} is ``How can I disable the web browser message in python'', however, our system output an ``opposite'' \textit{question title} of ``How to I open the web browser message when python2''. This example reveals that in some cases, \rev{question titles} can be generated incorrectly by only looking at the implementation details of the code snippet. This is because we can not judge the developers' intent just through the code snippet attached to the question.
    \item Also, \textbf{outputs from our system are not always ``perfect''}. The gap between ground truth \rev{question titles} and machine generated \rev{question titles} is still large. For example, in the last sample, The question quality of our model degrades on longer and compositional inputs. This indicates that there is still a large room for our question generation system to improve.
    It would be interesting to further investigate how to interpret why certain irrelevant words are generated in the \rev{question title}. For example, in the second and fifth samples, there are some irrelevant words at the end of generated questions. We will address such problems in the future.
\end{enumerate}
}

\rev{
\textbf{Answer to \textit{RQ-1:} How effective is our approach under automatic evaluation?} - we conclude that our approach is effective under automatic evaluation and beats the baselines by a large margin.
}

\subsection{\rev{\textbf{RQ-2: How effective is our approach compared with the CODE-NN model?}}}
\rev{
CODE-NN trained a neural attention model generate summaries of C\# and SQL code fragment, they have published their C\# and SQL datasets, which include 66,015 (title, query) pairs for C\# and 32,337 pairs for SQL. It is worth emphasizing that CODE-NN removed all the non-parsable code snippets and retained only the parsable code snippets. We retrained our approach on the CODE-NN datasets, the automatic evaluation results of our approach and CODE-NN model are summarized in Table~\ref{tab:codenn-eval}. Because CODE-NN use the BLEU-4 metric for evaluation, we only report the BLEU-4 score in our table. 
Apart from that, we also explored the effectiveness of transferring our trained model to the new datasets. We further applied the C\# and SQL model already obtained to the CODE-NN datasets. This is reasonable because CODE-NN extracted the code snippet only from the accepted answers containing exactly one code snippet, while our approach extracted the code snippet from the questions, so training dataset of our approach will not contaminate the CODE-NN datasets. In other words, our model does not see any test case in the CODE-NN dataset during the training process. From the table, we can observe the following points:
\begin{enumerate}
    \item In general, our approach and CODE-NN outperforms the other baselines by a large margin. The results are consistent with our previous evaluation. This further justifies the encoder-decoder architecture approach is helpful to learn the semantic and structural information from the code snippet.  
    \item The neural models, i.e., CODE-NN and ours, have better performance on C\# than SQL. This is probably due to the following reasons: First, generating question titles for SQL code snippets is a more challenging task since the SQL code snippet only has a handful of keywords and functions, and the generation models need to rely on other structural aspects. Second, the size of the SQL training data (32,337 pairs) is much smaller than the size of the C\# training data (66,015 pairs), it is more difficult to train a good neural model if there is lack of sufficient training data.
    \item By using CODE-NN datasets, our model performs better than CODE-NN. It improves BLEU-4 score by 7.8\% on C\# dataset and 10.8\% on SQL dataset. We attribute this to the \textit{copy} mechanism and \textit{coverage} mechanism incorporated into our approach, which is able to handle the low frequency tokens and reduce the redundancy during the generation process. 
    \item By transferring existing trained models to the CODE-NN datasets, it is notable that even without training directly on the CODE-NN datasets, we can still achieve comparable results compared with the CODE-NN model. We attribute this to the advantage of our model as well as the larger datasets constructed with our approach. We have collected more than 170K  $\langle$\textit{code snippet}, \textit{question}$\rangle$ pairs for C\# and more than 150K pairs for SQL. The CODE-NN datasets only include 60k+ C\# pairs and 30k+ SQL pairs. This verifies the importance of using big training data for applying deep learning-based methods in software engineering. 
\end{enumerate}
\textbf{Answer to \textit{RQ-2:} How effective is our approach compared with CODE-NN?} - we conclude that our approach is more effective compared with Code-NN.
}
\begin{table*}
\centering
\caption{\rev{Automatic evaluation(CODE-NN dataset)}}
\label{tab:codenn-eval}
\vspace*{-6pt}
\rev{
\resizebox{0.7\textwidth}{!}{
    \begin{tabular}{||l|c|c||}
    \hline
    Model & BLEU-4 (C\# Dataset) & BLEU-4 (SQL Dataset) \\
    \hline
    IR & $13.7$ & $13.5$ \\ \hline
    Moses & $11.6$ & $15.4$ \\ \hline
    CODE-NN & $20.5$ & $18.4$ \\ \hline
    Ours & $22.1$ & $20.4$ \\ \hline\hline
    Ours (Transfer) & $21.3$ & $18.4$ \\ 
    \hline
\end{tabular}
}
}
\end{table*}

\subsection{\textbf{RQ-3: How effective is our approach under human evaluation?}}
\subsubsection{\rev{Human Evaluation Results}}
\begin{figure}\vspace{-0.0cm}
\centerline{\includegraphics[width=0.73\textwidth]{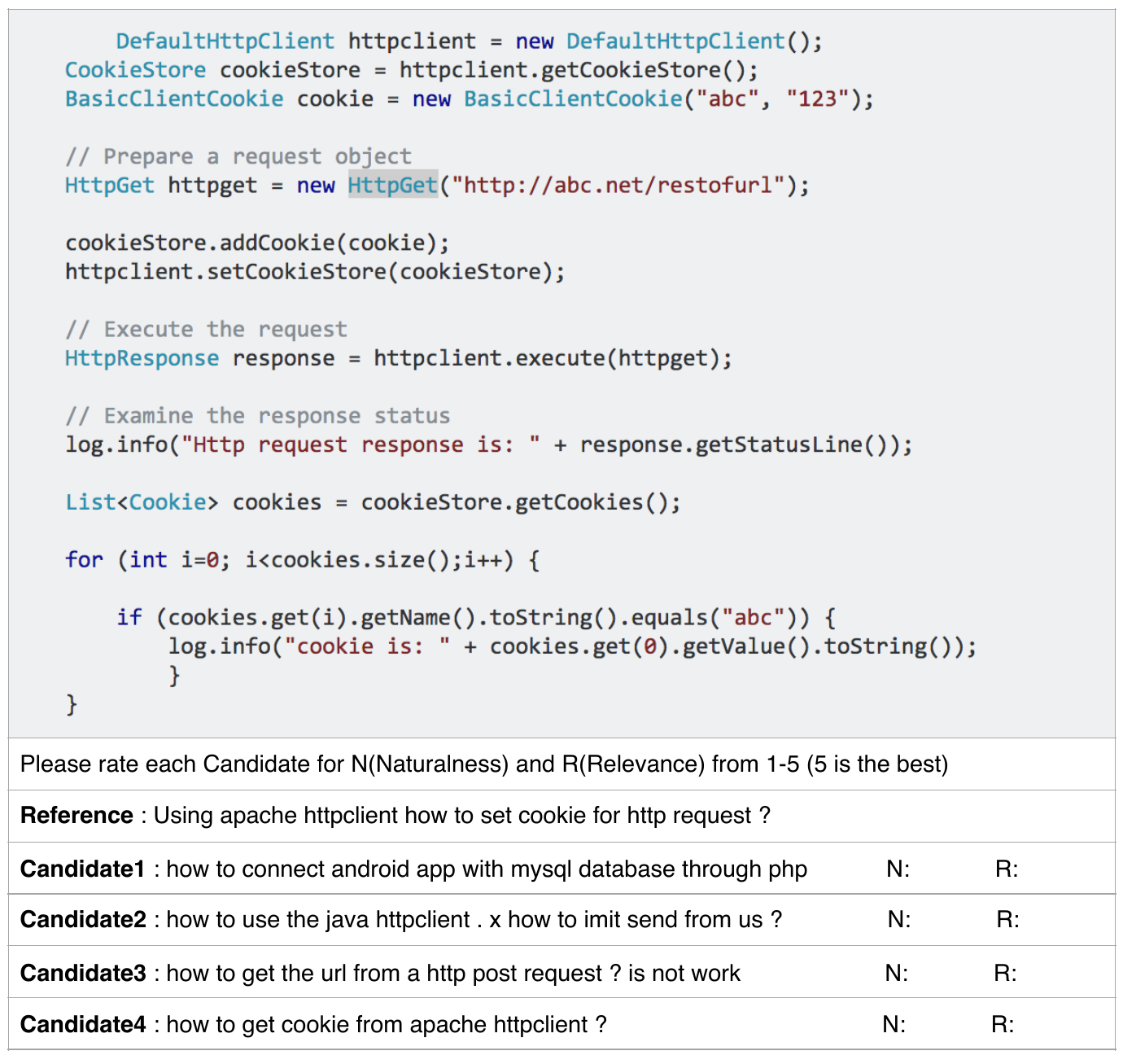}}
\vspace*{-0pt}
\caption{User Study Case (Human Evaluation)}
\label{fig:userstudy}\vspace{0.0cm}
\end{figure}

Fig.~\ref{fig:userstudy} shows one example in our \rev{human evaluation study}.
We obtain \rev{250} groups of scores from human evaluation for Python and Java Dataset respectively. Each group contains 4 pairs of scores, which were rated for candidates produced by IR, Moses, Seq2Seq and our approach. Each pair contains a score for the \textit{Naturalness} modality and a score for \textit{Relevance} modality. We regard a score of 1 and 2 as low-quality, a score of 3 as medium quality, and a score of 4 and 5 as high-quality.
Regarding human evaluation study results, the responses from all evaluators is then averaged for each modality. We also count the proportion of each quality type within each modality.
The quality distribution and average score of Naturalness and Relevance across each methods are presented in Table ~\ref{tab:humanpython} and Table ~\ref{tab:humanjava}.
From the table, several points stand out:
\begin{enumerate}
    \item From Naturalness prospective, \textbf{IR performs a slightly better than our approach}. This is reasonable since it retrieves other similar \rev{question titles} which are all also written by humans. However its output lacks the explanation to the actual input code snippet, which also explains its surprisingly low score on Relevance.

    \item From Relevance prospective, \textbf{the \rev{question titles} generated by our approach are much more appreciated} by the volunteers. Its superior performance in terms of Relevance further supports our claim that it manages to select content from input more effectively.

    \item In general, \textbf{our model performs well across both dimensions}. The results of human evaluation are consistent with automatic evaluation results. The considerable proportion of high-quality questions generated by our approach with respect to the \textit{Naturalness} and \textit{Relevance} also reconfirms the effectiveness of our system.
\end{enumerate}

\begin{table*} 
\centering
\caption{Human Evaluation(Python dataset)}
\label{tab:humanpython}
\vspace*{-6pt}
\rev{
\resizebox{0.95\textwidth}{!}{
    \vspace{-0.2cm}\begin{tabular}{||l| c c c c | c c c c ||}
        \hline
        \ Model  & Naturalness
                 & Low\textsubscript{N}
                 & Medium\textsubscript{N}
                 & High\textsubscript{N}
                 & Relevance
                 & Low\textsubscript{R}
                 & Medium\textsubscript{R}
                 & High\textsubscript{R} \\
        \hline
        \ IR & 3.91 & 13.2\% & 15.6\% & 71.2\% & 2.22 & 66.4\% & 19.2\% & 14.4\% \\
        \ Moses & 2.44 & 62.4\% & 17.2\% & 20.4\% & 2.73 & 40.8\% & 30.8\% & 28.4\% \\
        \ NMT & 3.38 & 22.0\% & 28.4\% & 49.6\% & 2.90 & 35.6\% & 32.4\% & 32.0\% \\
        \hline
        \ Ours & 3.75 & 18.4\% & 12.8\% & 68.8\% & 3.55 & 18.8\% & 22.8\% & 58.4\% \\
        \hline
    \end{tabular}
}
}
\end{table*}

\begin{table*} 
\centering
\caption{Human Evaluation(Java dataset)}
\label{tab:humanjava}
\vspace*{-6pt}
\rev{
\resizebox{0.95\textwidth}{!}{
    \vspace{-0.2cm}\begin{tabular}{||l|c c c c | c c c c ||}
        \hline
        \ Model  & Naturalness
                 & Low\textsubscript{N}
                 & Medium\textsubscript{N}
                 & High\textsubscript{N}
                 & Relevance
                 & Low\textsubscript{R}
                 & Medium\textsubscript{R}
                 & High\textsubscript{R} \\
        \hline
        \ IR & 3.56 & 19.6\% & 22.8\% & 57.6\% & 2.29 & 68.4\% & 14.4\% & 17.2\% \\
        \ Moses & 2.37 & 62.4\% & 18.4\% & 19.2\% & 2.24 & 65.2\% & 21.6\% & 13.2\% \\
        \ NMT & 2.96 & 28.0\% & 45.2\% & 26.8\% & 2.66 & 47.2\% & 27.6\% & 25.2\% \\
        \hline
        \ Ours & 3.42 & 22.0\% & 27.2\% & 50.8\% & 3.25 & 28.8\% & 24.4\% & 26.8\% \\
        \hline
    \end{tabular}
}
}
\end{table*}

\rev{
\textbf{Answer to \textit{RQ-3:} How effective is our approach under human evaluation?} 
In general,
for considering the combination of both modality, i.e., \textit{Naturalness} and \textit{Relevance}, our model beats the baselines by a large margin.
}

\subsection{\textbf{\rev{
RQ-4: How effective is our approach for improving low-quality questions?}
}}

\begin{figure}\vspace{-0.0cm}
\centerline{\includegraphics[width=0.85\textwidth]{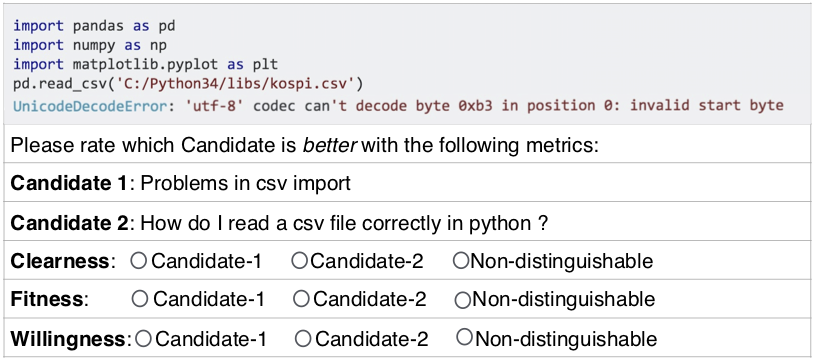}}
\vspace*{-5pt}
\caption{\rev{User Study Case (Practical Manual Evaluation)}}
\label{fig:userstudy_manual}\vspace{0.0cm}
\end{figure}

\subsubsection{\rev{Practical Manual Evaluation Results}}
\rev{
Fig.~\ref{fig:userstudy_manual} shows one example of our practical manual evaluation study. We collected 50 pairs of question titles (one was generated by humans and one was generated by our approach) for Python and Java respectively for comparison purposes. For each pairwise comparison, we got 5 groups of selections from the evaluators. Each group contains three user selections with respect to the \textit{Clearness}, \textit{Fitness} and \textit{Willingness} measures respectively. We calculated the proportion of the user selection according to each evaluation metric. Table~\ref{tab:manual_evaluation_python} and Table~\ref{tab:manual_evaluation_java} show the results of the practical manual evaluation for Python and Java respectively. From the table we can see that:
\begin{enumerate}
    \item The question titles generated by our approach outperform the poor quality question titles in terms of all the metrics. This demonstrates that our approach produces more clear and/or appropriate question titles, which is potentially helpful for improving the low-quality questions in Stack Overflow. 
    \item Particularly, our question titles have substantially better willingness scores, indicating that developers are more willing to respond to our questions. 
    This shows that question titles generated by our model are more likely to elicit further interactions, which is helpful to increase the likelihood of receiving answers.
\end{enumerate}
}

\begin{table} \vspace{-0.0cm}
\rev{
\caption{\rev{Practical Manual Evaluation (Python dataset)}}
\begin{center}
\vspace{-0.2cm}\begin{tabular}{|| c | c | c | c ||}
    \hline
    \ Ours vs. Human & Win (\%) & Lose (\%) & Non-distinguishable (\%) \\
    \hline\hline
    Clearness & 52.4 & 33.2 & 14.4 \\
    \hline
    Fitness   & 55.2 & 24.0 & 20.8  \\
    \hline
    Willingness  & 61.2 & 31.6 & 7.2 \\
    \hline
\end{tabular}
\label{tab:manual_evaluation_python}
\end{center}
}
\vspace{-0.0cm}
\end{table}

\begin{table} \vspace{-0.0cm}
\rev{
\caption{\rev{Practical Manual Evaluation (Java dataset)}}
\begin{center}
\vspace{-0.2cm}\begin{tabular}{|| c | c | c | c ||}
    \hline
    \ Ours vs. Human & Win (\%) & Lose (\%) & Non-distinguishable (\%) \\
    \hline\hline
    Clearness & 42.8 & 34.0 & 23.2  \\
    \hline
    Fitness   & 47.2 & 39.6 & 13.2 \\
    \hline
    Willingness   & 49.2 & 26.8 & 24.0  \\
    \hline
\end{tabular}
\label{tab:manual_evaluation_java}
\end{center}
}
\vspace{-0.0cm}
\end{table}

\subsubsection{\rev{Examples of Practical Manual Evaluation}}
\begin{figure}\vspace{-0.0cm}
\centerline{\includegraphics[width=0.95\textwidth]{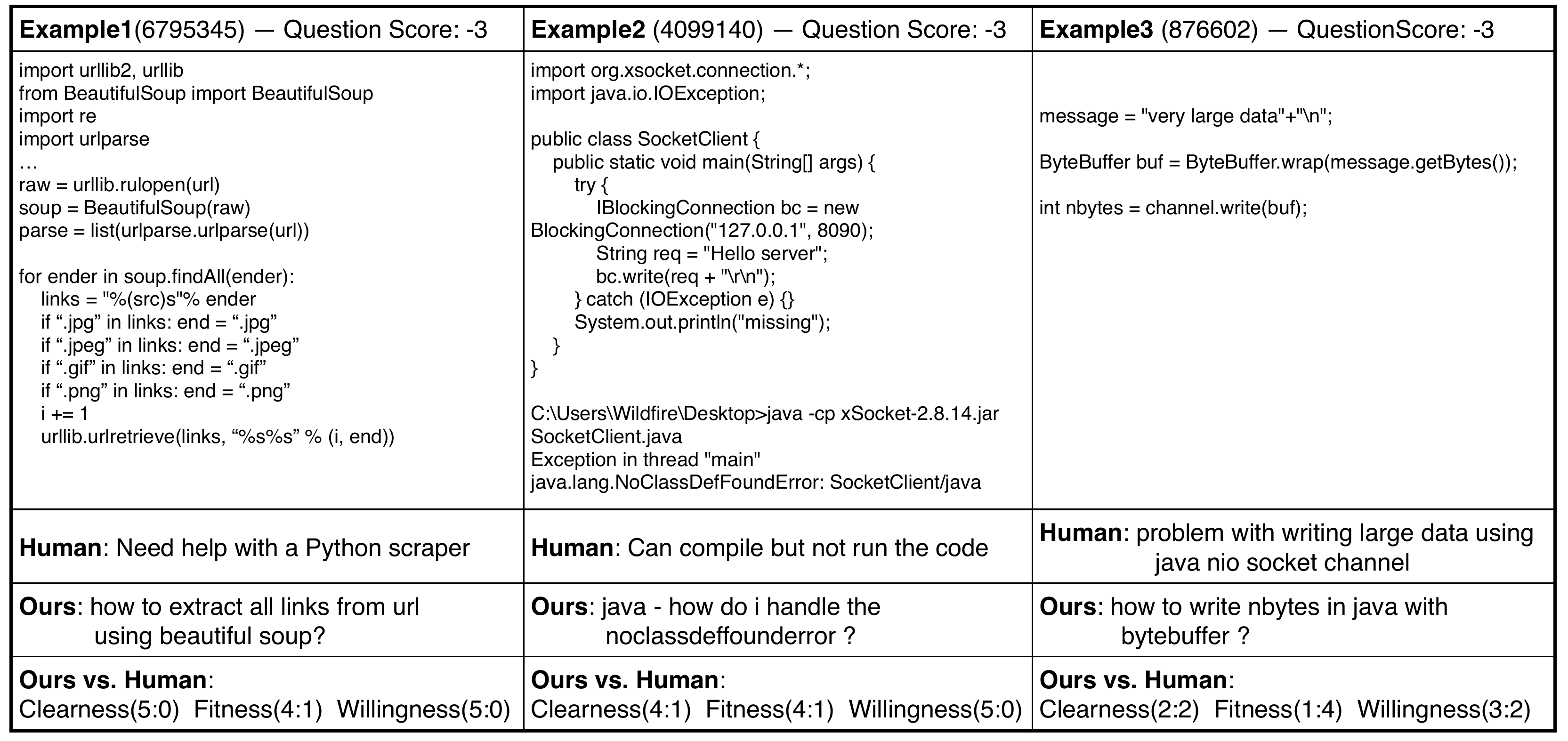}}
\vspace*{-0pt}
\caption{\rev{Practical Manual Evaluation Example}}
\label{fig:manual_example}\vspace{0.0cm}
\end{figure}

\rev{
Fig.~\ref{fig:manual_example} presents three examples of manual evaluation results. From these cases we can see that: 
\begin{enumerate}
    \item The question titles with poor scores in Stack Overflow are often unclear (e.g., Example1) and/or unappropriate (e.g., Example2). For such cases, the question titles generated by our approach are more clear and attractive, such as Example1, and also questioning on key information. For example, the newly generated question titles in Example2 are much more appreciated by the evaluators than the original ones, which increases the likelihood and willingness of the developers to offer help.
    \item Not all of the poor quality question titles can be improved by our approach. Notably for some posts, our approach suffered from semantic drift, that is the questions generated by our approach do not align well with the developers' intent. Such as in Example3, the developer's problem was more about ``writing with large data'', while the semantics of our question generated has drifted to the problem of ``java with bytebuffer''. This is because the string variable ``very large data'' has been replaced by STR during data preprocessing, such information loss hinders the learning process of our approach.
    \item Even though the results generated by our approach are still not perfect, our approach is the first step on this topic and we also release our code and dataset to inspire further follow-up work. 
\end{enumerate}
}
\rev{
\textbf{Answer to \textit{RQ-4:} How effective is our approach for improving low-quality questions?} 
In general, for a large number of low-quality questions in Stack Overflow, our approach can improve the quality of the question titles via \textit{Clearness}, \textit{Fitness} and \textit{Willingness} measures.
}

\subsection{\textbf{\rev{RQ-5: How effective is our use of \textit{attention} mechanism, \textit{copy} mechanism and \textit{coverage} mechanism under automatic evaluation?}}}
\subsubsection{\rev{Ablation Analysis Results}}
We added an \textit{attention} mechanism, a \textit{copy} mechanism and a \textit{coverage} mechanism to our sequence-to-sequence architecture. The ablation analysis is to verify the effectiveness of the three mechanisms, to be more specific, we compare our approach with several of its incomplete variants:
\begin{itemize}
    \item \textbf{Model\textsubscript{Atten+Copy}} removes the \textit{coverage} mechanism from our approach.
    \item \textbf{Model\textsubscript{Atten+Coverage}} removes the \textit{copy} mechanism from our approach.
    \item \textbf{Model\textsubscript{Atten}} removes the \textit{copy} and \textit{coverage} mechanism from our approach.
    \item \textbf{Model\textsubscript{Basic}} removes all the \textit{attention}, \textit{copy} and \textit{coverage} mechanism from our approach.
\end{itemize}
The ablation analysis results are presented in the Table ~\ref{tab:pythonablation} and Table ~\ref{tab:javaablation}. We can observe the following points:

\begin{enumerate}
    \item By comparing the results of \textbf{Model\textsubscript{Basic}} and \textbf{Model\textsubscript{Atten}}, it is clear that incorporating an \textit{attention} mechanism is able to improve the overall performance. For example, by adding \textit{attention} mechanism, \rev{the average BLEU-4 score of the Attention-based model was improved by 9\% and 13.3\%, ROUGE-L score was improved by 6.8\% and 10.8\% on Python and Java dataset respectively.} We attribute this to the ability of \textit{attention} mechanism to perform better content selection, which can focus on the more salient part of the source code snippet.
    \item By comparing \textbf{Model\textsubscript{Atten}} with \textbf{Model\textsubscript{Atten+Copy}} and \textbf{Model\textsubscript{Atten+Coverage}}, we can measure the performance improvements achieved due to the incorporation of \textit{copy} mechanism and \textit{coverage} mechanism respectively. Better performance can be achieved by solely adding \textit{copy} or \textit{coverage} mechanism to the attention-based model. This signals that both \textit{copy} and \textit{coverage} mechanism do have contributions to the performance improvements.
    \item Without \textit{copy} mechanism, \rev{there is a drop overall in every evaluation measure, the ROUGE-L score drops 13\% and 9.4\% on Python and Java dataset respectively}. On the other hand, without \textit{coverage} mechanism, we see a consistent but sufficiently lower drop in each evaluation measure, \rev{the ROUGE-L drops 12.3\% on Python and 3.8\% on Java.}
    \item By comparing the results of our approach with each of the variant model, we can see that no matter which type of mechanism we dropped, it does hurt the performance of our model. This verifies the importance and effectiveness of these three mechanisms.
\end{enumerate}
\subsubsection{\rev{Examples of Ablation Analysis}}

\begin{figure}\vspace{-0.0cm}
\centerline{\includegraphics[width=0.85\textwidth]{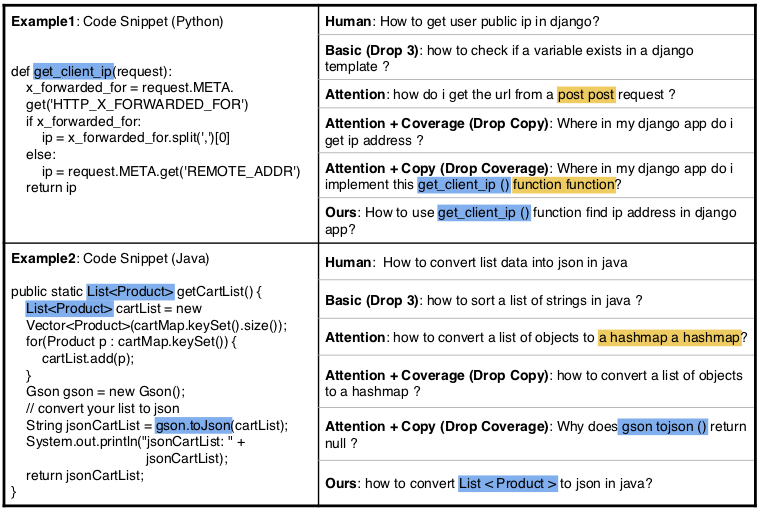}}
\vspace*{-0pt}
\caption{\rev{Ablation Analysis Example}}
\label{fig:ablation_example}\vspace{0.0cm}
\end{figure}

\rev{
To gain further insight into our approach, we further illustrate some examples from the ablation analysis to show the effect of employing the \textit{attention}, \textit{copy} and \textit{coverage} mechanism. The results are shown in Fig.~\ref{fig:ablation_example}, we can see that:
\begin{enumerate}
     \item \rev{Question titles} generated by the basic model are of low-quality. Comparing the results of the basic model and attention model, we can see that by adding the \textit{attention} mechanism, the generated question titles are more meaningful and relevant for the given code snippet. The \textit{attention} mechanism enables the model to focus on the relevant parts of the input sequence as needed. As shown in Example1, the model will focus on the ``request'' related segment in source code when it generates ``post request'' for the question title.  
     \item Repetition is a common problem for attentional sequence to sequence models (e.g.,~\cite{tu2016modeling, sankaran2016temporal, suzuki2016rnn}). Meaningless repeated words are produced during the generation process (highlighted with yellow color). 
     We introduce a \textit{coverage} mechanism for discouraging such repetitions in our generator by quantitatively emphasizing the coverage of sentence words while decoding. As can be seen in Example2, ``a harshmap'' has been meaningless repeated twice, employing the \textit{coverage} mechanism can effectively discourage such repetitions. 
    \item We observe that a \textit{high-quality} question title is generated using our approach. Recall that a code snippet usually contains tokens (highlighted with a blue color) with very rare occurrences. It is difficult for a decoder to generate such words solely based on language modeling. For such cases, we incorporate the \textit{copy} mechanism to copy the rare tokens from the code snippet to the question title. In the first example, the method name \textit{get\_client\_ip} has been properly picked up from the source code snippet to the generated question titles. 
\end{enumerate}
}
\rev{
\textbf{Answer to \textit{RQ-5:} How effective is our use of \textit{attention} mechanism, \textit{copy} mechanism and \textit{coverage} mechanism under automatic evaluation?}
In summary, all the three mechanisms, i.e., \textit{attention} mechanism, \textit{copy} mechanism, \textit{coverage} mechanism, are effective and helpful to enhance the performance of our approach.
}

\begin{table*}[htb]
\centering
\caption{Ablation evaluation (Python dataset)}
\label{tab:pythonablation}
\vspace*{-6pt}
\rev{
\resizebox{0.95\textwidth}{!}{
    \begin{tabular}{||l|ccccc||}
      \hline
      Measure
      & Model\textsubscript{Basic}
      & Model\textsubscript{Atten}
      & Model\textsubscript{Atten+Coverage}
      & Model\textsubscript{Atten+Copy}
      & Ours \\
      \hline
      BLEU-1 & $25.1\pm1.5\%$
             & $28.6\pm1.7\%$
             & $29.6\pm1.8\%$
             & $31.5\pm1.9\%$
             & $35.8\pm2.0\%$ \\
      BLEU-2 & $20.2\pm0.7\%$
             & $22.3\pm0.8\%$
             & $24.6\pm0.6\%$
             & $27.8\pm0.8\%$
             & $30.1\pm0.9\%$ \\
      BLEU-3 & $19.1\pm0.4\%$
             & $21.7\pm0.4\%$
             & $23.8\pm0.5\%$
             & $25.4\pm0.4\%$
             & $26.8\pm0.4\%$ \\
      BLEU-4 & $18.7\pm0.3\%$
             & $20.3\pm0.3\%$
             & $22.3\pm0.2\%$
             & $23.1\pm0.2\%$
             & $24.2\pm0.3\%$ \\
      ROUGE-1 & $32.8\pm2.0\%$
             & $34.1\pm2.3\%$
             & $35.3\pm2.2\%$
             & $35.4\pm2.4\%$
             & $39.9\pm2.5\%$ \\
      ROUGE-2 & $9.1\pm0.8\%$
             & $10.2\pm1.2\%$
             & $10.6\pm2.1\%$
             & $10.8\pm2.0\%$
             & $12.6\pm2.5\%$ \\
      ROUGE-L & $29.2\pm5.8\%$
             & $31.2\pm2.0\%$
             & $31.9\pm2.1\%$
             & $32.2\pm2.2\%$
             & $36.7\pm2.4\%$ \\
      \hline
\end{tabular}
}
}
\end{table*}

\begin{table*}[htb]
\centering
\caption{Ablation evaluation (Java dataset)}
\label{tab:javaablation}
\vspace*{-6pt}
\rev{
\resizebox{0.95\textwidth}{!}{
    \begin{tabular}{||l|ccccc||}
      \hline
      Measure
      & Model\textsubscript{Basic}
      & Model\textsubscript{Atten}
      & Model\textsubscript{Atten+Coverage}
      & Model\textsubscript{Atten+Copy}
      & Ours \\
      \hline
      BLEU-1 & $20.5\pm1.0\%$
             & $25.2\pm1.6\%$
             & $27.8\pm1.6\%$
             & $29.7\pm1.7\%$
             & $31.8\pm1.8\%$ \\
      BLEU-2 & $16.4\pm0.6\%$
             & $20.7\pm0.7\%$
             & $25.0\pm0.6\%$
             & $26.1\pm0.6\%$
             & $27.5\pm0.7\%$ \\
      BLEU-3 & $17.8\pm0.4\%$
             & $21.1\pm0.3\%$
             & $23.6\pm0.3\%$
             & $24.4\pm0.3\%$
             & $25.2\pm0.3\%$ \\
      BLEU-4 & $18.1\pm0.2\%$
             & $20.5\pm0.2\%$
             & $22.0\pm0.1\%$
             & $22.6\pm0.2\%$
             & $23.3\pm0.2\%$ \\
      ROUGE-1 & $28.3\pm1.3\%$
             & $30.5\pm2.0\%$
             & $31.2\pm2.0\%$
             & $33.2\pm2.1\%$
             & $35.4\pm2.2\%$ \\
      ROUGE-2 & $6.9\pm0.5\%$
             & $7.9\pm1.1\%$
             & $8.2\pm1.2\%$
             & $8.7\pm1.5\%$
             & $10.0\pm1.8\%$ \\
      ROUGE-L & $24.6\pm1.1\%$
             & $27.3\pm1.8\%$
             & $28.8\pm1.9\%$
             & $30.6\pm2.0\%$
             & $31.8\pm2.2\%$ \\
      \hline
\end{tabular}
}
}
\end{table*}

\subsection{\textbf{\rev{RQ-6: How effective is our approach under different parameter settings?}}}
One of the key parameter of our approach is the vocabulary size. The encoder-decoder architecture models need a fixed vocabulary for the source input and target output. To generate all the possible words, the basic Seq2Seq model has to include all the vocabulary tokens that appeared in the training set, which requires a lot of time and memory to train the models. One advantage of our model is that, with the help of \textit{copy} mechanism, our approach can copy words from source input to the target output. We can maintain a small size vocabulary which exclude the low frequency words, but also get better performance and generalization ability.

\begin{table} \vspace{-0.0cm}
\caption{Vocab Size \& Training Time(per epoch)}
\begin{center}
\vspace{-0.4cm}
\begin{tabular}{||c|c|c|c||}
    \hline
    \multirow{8}{*}{Python}   & Threshold  & Vocab Size & Training Time \\ \cline{2-4}
                              & 1  & 58,536 & 766.9 \\ \cline{2-4}
                              & 2  & 49,656 & 719.1 \\ \cline{2-4}
                              & 3  & 36,277 & 663.7 \\ \cline{2-4}
                              & 5  & 22,244 & 593.8 \\ \cline{2-4}
                              & 7  & 16,368 & 549.2 \\ \cline{2-4}
                              & 10  & 12,142 & 539.1 \\ \cline{2-4}
                              & 100 & 2,503  & 499.9 \\ \cline{2-4}
    \hline\hline
    \multirow{8}{*}{Java}   & Threshold  & Vocab Size & Training Time \\ \cline{2-4}
                          & 1  & 221,160 & 2218.3 \\ \cline{2-4}
                          & 2  & 131,862 & 1692.3 \\ \cline{2-4}
                          & 3  & 79,048  & 1074.1 \\ \cline{2-4}
                          & 5  & 54,352  & 962.2 \\ \cline{2-4}
                          & 7  & 38,670  & 898.4 \\ \cline{2-4}
                          & 10 & 27,341  & 831.4 \\ \cline{2-4}
                          & 100 & 4,642    & 723.8 \\ \cline{2-4}

    \hline
\end{tabular}
\label{tab:vocab_size}
\end{center}
\vspace{-0.0cm}
\end{table}

\begin{figure}\vspace{-0.0cm}
\centerline{\includegraphics[width=0.45\textwidth]{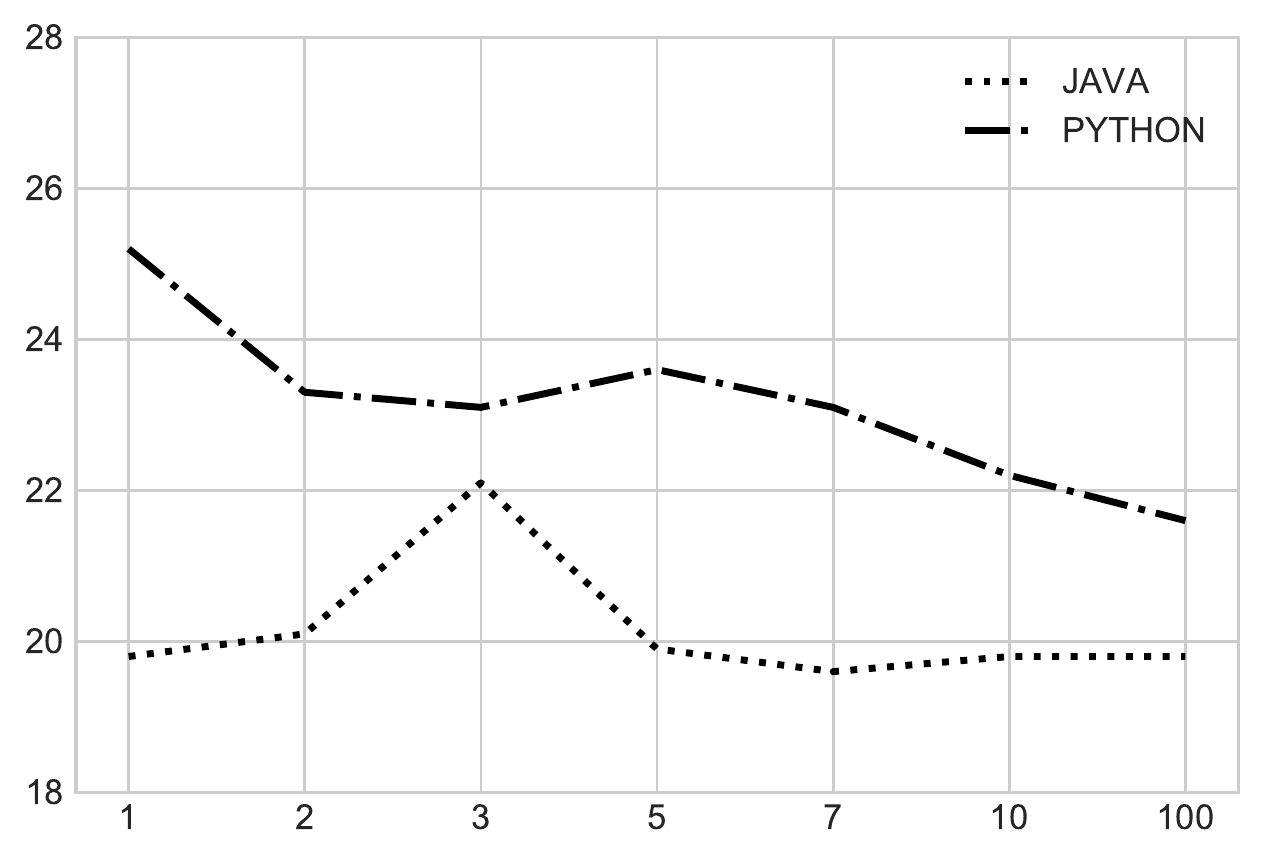}}
\vspace*{-2pt}
\caption{BLEU4 Score under different vocab threshold}
\label{fig:vocab-bleu-tuning}\vspace{0.1cm}
\end{figure}

\begin{figure}\vspace{-0.0cm}
\centerline{\includegraphics[width=0.52\textwidth]{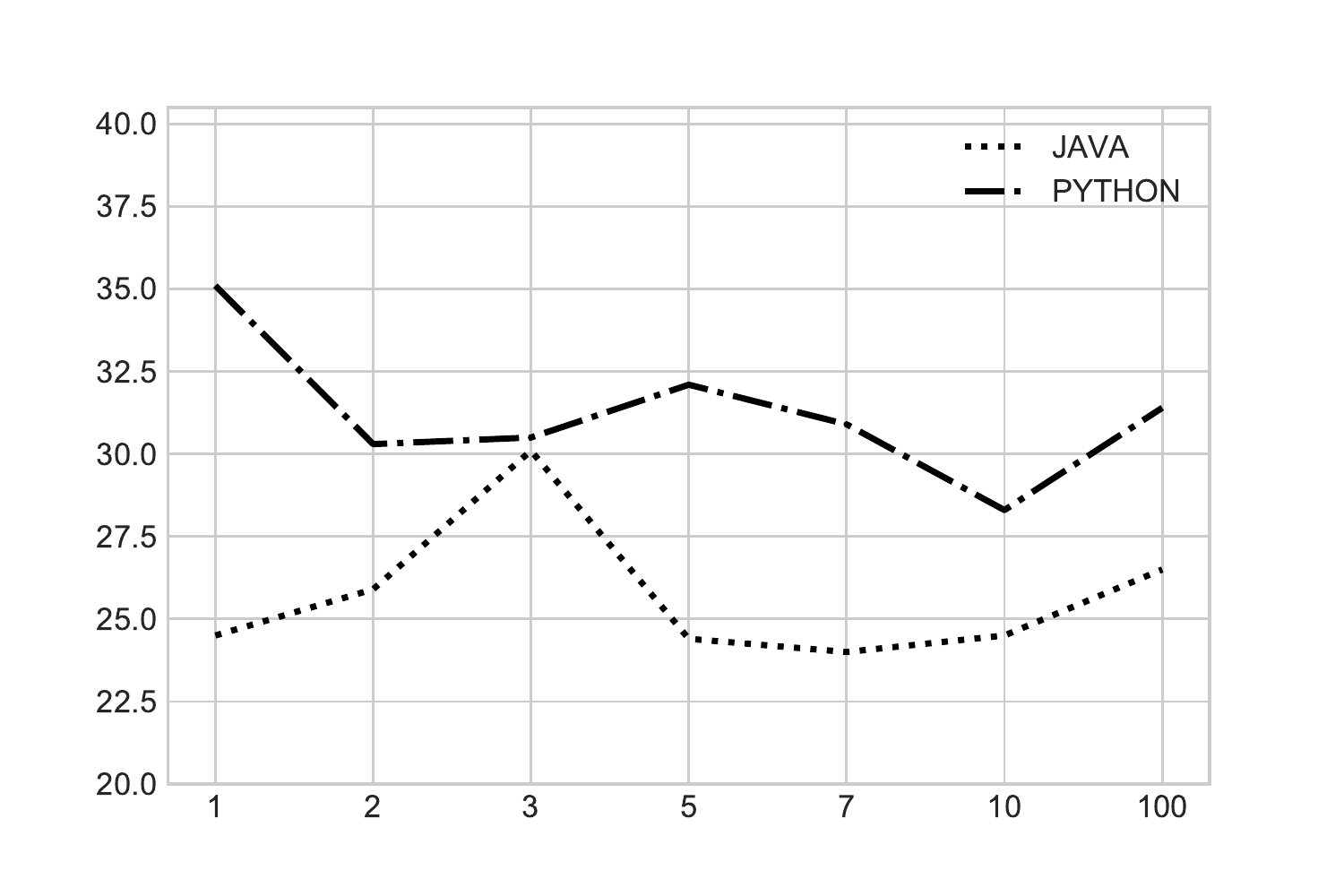}}
\vspace*{-2pt}
\caption{ROUGE-L Score under different vocab threshold}
\label{fig:vocab-rouge-tuning}\vspace{-0.0cm}
\end{figure}

We set different word frequency threshold, i.e., 1, 2, 3, 5, 7, 10, 100, for constructing the vocabulary. Setting word frequency threshold to 1 means the vocabulary is constructed with words that appeared at least twice in the training set. Different models were trained under these parameters on the Python and Java datasets separately. The vocabulary size and training time under different threshold are summarised in Table~\ref{tab:vocab_size}. Fig.~\ref{fig:vocab-bleu-tuning} and Fig.~\ref{fig:vocab-rouge-tuning} shows the influence of different threshold settings on the BLEU-4 score and ROUGE-L score. We have the following observations from these figures:
\begin{enumerate}
    \item Our approach achieves its best performance on Java dataset when the similarity threshold set to 3, the corresponding vocabulary size is 79,048. When the vocabulary size is too big, i.e., 221,160 with threshold equals 1, the BLEU4 and ROUGE-L score becomes lower.
    This is because some non-generic words will be included in the fixed vocabulary, which leads to difficulties for our approach to learn how to copy words from the input source sequence.
    \item The results of our approach are best on Python dataset when the word frequency threshold set to 1, the corresponding vocabulary size is 58,536. Compared with the results of the Java dataset, the optimum vocabulary size settings of our approach can be around 60000.
    \item When the word frequency threshold rockets up to 100, the vocabulary size decreases to 2,503 and 4,626 on Python and Java dataset respectively. Even with a much smaller vocabulary size, our approach can still have a comparable performance against Basic Seq2Seq model, which further supports the generalization ability of our approach.
\end{enumerate}

\begin{figure}\vspace{-0.0cm}
\centerline{\includegraphics[width=0.45\textwidth]{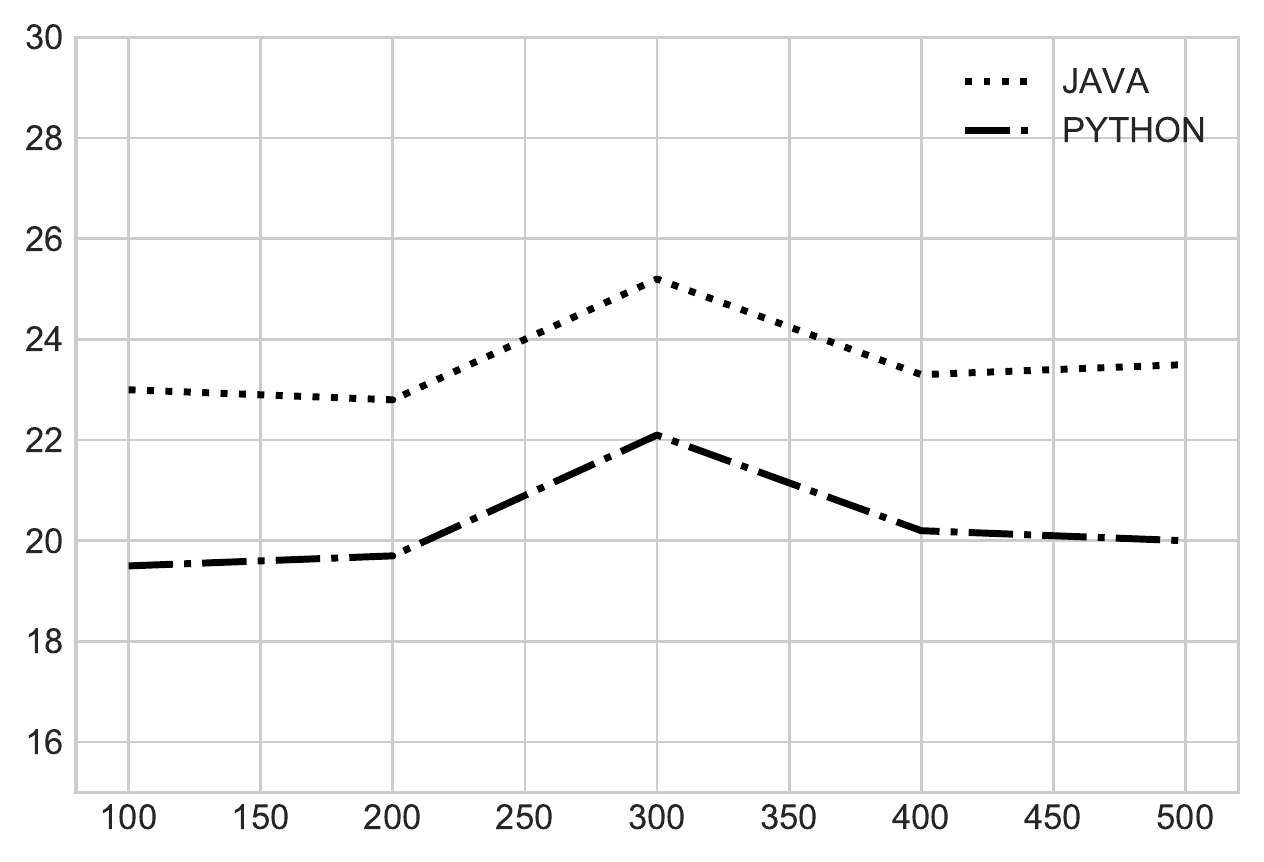}}
\vspace*{-5pt}
\caption{BLEU4 Score under different sizes of word embeddings}
\label{fig:dim-bleu-tuning}\vspace{0.1cm}
\end{figure}

\begin{figure}\vspace{-0.0cm}
\centerline{\includegraphics[width=0.45\textwidth]{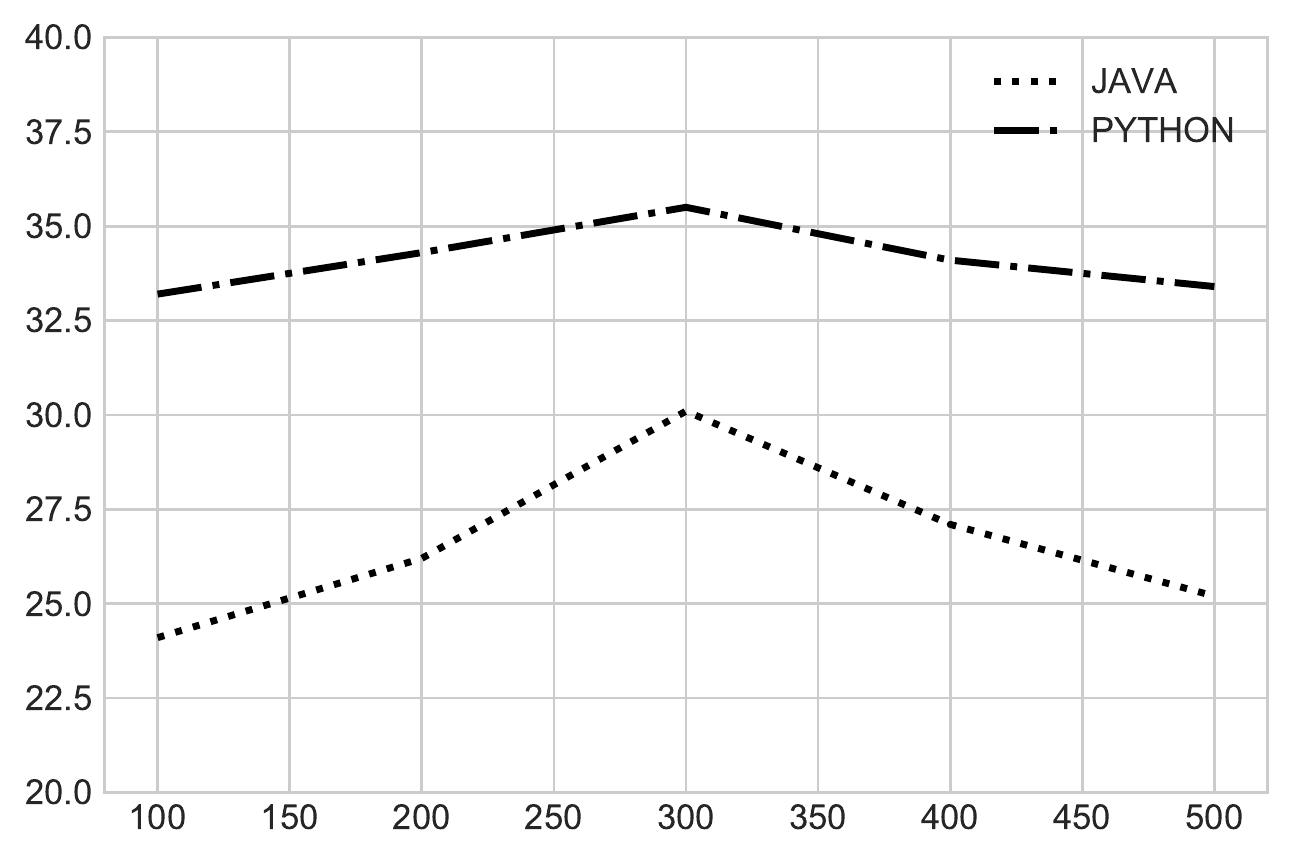}}
\vspace*{-5pt}
\caption{ROUGE-L Score under different sizes of word embeddings}
\label{fig:dim-rouge-tuning}\vspace{-0.0cm}
\end{figure}

Another parameter of our approach is the dimension of word embeddings. We choose five different word embedding sizes, i.e., 100, 200, 300, 400, 500, and qualitatively compare the performance of our approach in these different word embeddings. Fig.~\ref{fig:dim-bleu-tuning} and Fig.~\ref{fig:dim-rouge-tuning} show  the influence of different word embedding sizes on the BLEU-4 and ROUGE-L score. One can clearly see that our approach achieves the best BLEU-4 and ROUGE-L score when the embedding size is set to 300. Too large word embedding size may not be helpful to improve the accuracy.

\subsection{\rev{RQ-7: How efficient is our approach in practical usage?}}
\rev{
The experiment was conducted on an Nvidia GeForce GTX 1080 GPU with 8GB memory. 
The time cost of our approach is mostly for the training process which takes approximately 8 to 10 hours for different datasets. The testing process on around 3,000 examples takes one to three minutes, while generating a single \rev{question title} only costs 20 to 60ms.  
}

\begin{figure}\vspace{-0.0cm}
\centerline{\includegraphics[width=0.85\textwidth]{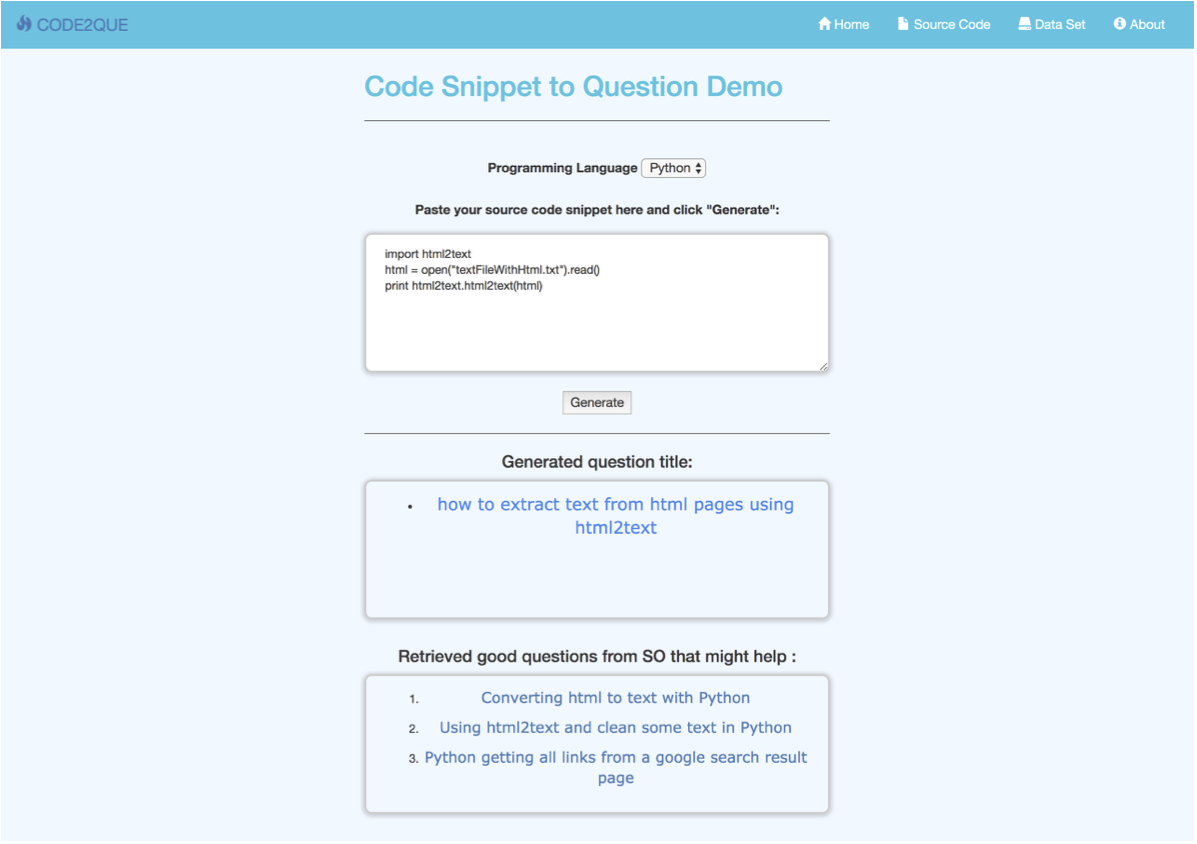}}
\vspace*{0pt}
\caption{CODE2QUE Web Service Tool}
\label{fig:web_tool}\vspace{-0.0cm}
\end{figure}

\rev{
Considering that the query for generating a question title using our approach is efficient, we have implemented our approach as a standalone web-based tool, named {\sc Code2Que}, to facilitate developers in using our approach and to inspire follow up research. Fig.~\ref{fig:web_tool} shows the web interface of {\sc Code2Que}. Developers can copy and paste their code snippet into our web application. {\sc Code2Que} embeds the code snippet via source code encoder and generates the question titles for the developers. We below describe the details of the input and output of such a process. 
\begin{itemize}
    \item \textbf{Input:} the input to the {\sc Code2Que} is a code snippet, which is an ordered sequence of source code lines. We have provided support for different types of programming languages (e.g., Python, Java, Javascript, C\# and SQL) for users to select.
    The input box in Fig.~\ref{fig:web_tool} shows an example of a Python code snippet. After inputting the code snippet, the developers can click the ``Generate'' button to submit their query. 
    \item \textbf{Output:} the output of {\sc Code2Que} is in two parts: (i) Generated Questions: {\sc Code2Que} will generate a question title using our backend model according to the code snippet and programming language they choose. For example, ``\textit{how to extract text from html pages using html2text}'' is generated for the given code snippet.  (ii) Retrieved Questions: After the developer submits his/her code snippet to the server, the code snippet is converted into a vector by our backend Source Code Encoder, then {\sc Code2Que} searches through our codebase and returns the top3 questions with similar code snippets. The link to these questions on the Stack Overflow website is also provided for reference. Developers can use these to quickly browse the related questions to have a better understanding of the problem.   
\end{itemize}
}

\rev{
\textbf{Answer to \textit{RQ-7:} How efficient is our approach in practical usage?} 
In summary, our approach is efficient enough for practical use and we have implemented a web service tool, named {\sc Code2Que}, to apply our approach for practical use. 
}

\section{Discussion}
\label{sec:discussion}
\revv{In this section, we discuss the main contribution of our work and analyze the strength and potential weakness of our work associated with each contribution.}
\subsection{Question Quality Improvement}
\revv{
It is important for CQA forums to maintain a satisfactory quality level for the questions and answers so as to improve community reputation and provide better user experience.
Questions are a fundamental aspect of a CQA website. Poorly formulated questions are less likely to receive useful responses, thus hindering the overall knowledge generation and sharing process.
\begin{itemize}
    \item \textit{Strength of our work}. Previous work related to CQA quality studies focus on question quality prediction. For example, the authors in \cite{ravi2014great} developed a model for predicting question quality using the content of the question. The authors in \cite{arora2015good} proposed a method to identify inappropriate questions by using previously asked similar questions. Different from the existing research, our study aims to improve low-quality questions in Stack Overflow. To the best of our knowledge, this is the first work that investigates the possibility of automatically improving low-quality questions in Stack Overflow.
    \item \textit{Weakness of our work}. According to our practical manual evaluation results, our approach can improve a large number of low-quality questions in Stack Overflow via \textit{Clearness}, \textit{Fitness} and \textit{Willingness} measures. However, the results generated by our approach are still not perfect, and for some posts, our approach suffers from semantic drift problems. We plan to incorporate more context information for generating better question titles in the future.
\end{itemize}
}

\subsection{Deep Sequence to Sequence Approach}
\revv{
Recently, deep learning has achieved promising results in solving many software engineering tasks, such as code search (e.g., ~\cite{gu2018deep, li2019neural, husain2019codesearchnet}), code summarization (e.g., ~\cite{iyer2016summarizing, hu2018deep, jiang2017automatically, wan2018improving}), and API recommendation (e.g., ~\cite{gu2016deep,  gu2017deepam}).
Among these works, a number of researchers have applied the sequence to sequence methods for mining the $\langle$natural language, code snippet$\rangle$ pairs, such as the commit message generation. (e.g., \cite{iyer2016summarizing, jiang2017automatically}).
\begin{itemize}
    \item \textit{Strength of our work}. A major challenge for question generation tasks in our study is the semantic gap between the code snippet and natural language descriptions. To bridge the gap between code fragment and natural language queries, we employed a deep sequence to sequence approach to build the neural language model for both code snippets and natural language questions. The neural language model automatically learns common patterns from the large scale source code snippets. Furthermore, different from the existing sequence to sequence learning approach, we add \textit{attention}, \textit{copy} and \textit{coverage} mechanism to our sequence-to-sequence architecture to suit our specific task. The \textit{attention} mechanism can perform better content selection from the input, while the \textit{copy} mechanism can handle the rare word problems among the code snippet, and the \textit{coverage} mechanism can eliminate the meaningless repetitions.
    \item \textit{Weakness of our work}. Previous works \cite{iyer2016summarizing, hu2018deep, wan2018improving} have shown that incorporating structural information of the source code (i.e., the AST) can improve the performance of the model, However, considering that the majority of the code snippets are not parsable in Stack Overflow, we do not use the AST structural information at the current stage. We plan to use the program repair algorithm to fix the code snippet in Stack Overflow and employ more contextual information of the source code in the future.    
\end{itemize}
}

\subsection{Question Generation Task}
\revv{
Stack Overflow is a collaborative question answering website, its target audience are software developers, maintenance professionals and programmers. Over the recent years, Stack Overflow has attracted increasing attention from the software engineering research community. However, since the questions and answers posted by developers on Stack Overflow are usually unstructured natural language texts containing code snippets, which makes it more challenging for researchers to mine and analyze these posts.
\begin{itemize}
    \item \textit{Strength of our work}. To improve the software development process, researchers have investigated the Stack Overflow knowledge-base for various software development activities, such as predicting the post quality ~\cite{ravi2014great, yao2013want, yang2014asking, arora2015good, ponzanelli2014understanding}, answer recommendation ~\cite{gkotsis2014s, xu2017answerbot, singh2016using}, code/questions retrieval ~\cite{xu2018domain, chen2016learning, allamanis2015bimodal, ganguly2015partially, henbeta2012semi} etc. However, to the best of our knowledge, this is the first work which investigates the question generation task in Stack Overflow. We first perform such a task to assist developers to generate a question title when presenting a code snippet.
    \item \textit{Weakness of our work}. We collected more than 1M $\langle$\textit{code snippet, \textit{question}}$\rangle$ pairs from Stack Overflow, which covers a variety of programming languages (e.g., Python, Java, Javascript, C\# and SQL). Considering our study is the first step on this topic, we have published our data to inspire further follow-up work. However, even though we have cleaned the data via pre-processing, some data may still be noisy. We plan to improve the dataset quality by further manual checking in the future.
\end{itemize}
}

\section{Threats to Validity}
\label{sec:threats}
\rev{
We have identified the following threats to validity among our study:
}

\vspace{0.1cm}\noindent\textbf{Internal Validity}
\rev{
Threats to internal validity are concerned
with potential errors in our code implementation and study settings. For the automatic evaluation, in order to reduce errors, we have double-checked and fully tested our source code. We have carefully tuned the parameters of the baseline approaches and used them in their highest performing settings for comparison, but there may still exist errors that we did not note. Considering such cases, we have published our source code and dataset to facilitate other researchers to replicate and extend our work.
}

\vspace{0.1cm}\noindent\textbf{External Validity}
\rev{
The external validity relates to the quality and generalizability of our dataset. Our dataset is constructed from the official Stack Overflow data dump which contains a variety of programming languages, such as Python, Java, Javascript, C\# and SQL. However, there are still many other programming languages in Stack Overflow which are not considered in our study. We believe that our results will generalize to other programming languages, due to the overall reasonable similarity in code snippets despite particular language syntax, semantics and APIs. We will try to extend our approach to other programming languages to benefit more users in future studies. 
}

\vspace{0.1cm}\noindent\textbf{Construct Validity}
\rev{
The construct validity concerns the relation between theory and observation. In this study, such threats are mainly due to the suitability of our evaluation measures. For the practical manual evaluation, the manual validation could be affected by the subjectiveness of the evaluators and the human errors.
For the human evaluation, the evaluators' degree of carefulness, effort and English skills in the examination process may affect the validity of judgements. We minimized such threats by choosing experienced participants who have at least one year of studying/working experience in English speaking countries, and are familiar with Python and Java programming languages. We also gave the participants enough time to complete the evaluation tasks. 
}

\vspace{0.1cm}\noindent\textbf{Conclusion Validity}
\rev{
The conclusion validity relates to issues that could affect the ability to draw correct conclusions about relations between the treatment and the outcome of an experiment.
One issue during the data filtering procedure is that we only keep the questions which contain several keywords, such as ``how'', ``what'', ``why''. However, since the questions in Stack Overflow can be rather complicated, our results do not shed light on how effective our solution is on questions of other kinds. 
On the other hand, from the human evaluation analysis, we see a key challenge for our current work is that the questions generated by our approach suffered from semantic drift. This is because it is difficult to judge a question poster's intent by solely looking at his/her code snippet. In such a case, more relevant information such as question description, question tags could further be incorporated within our model, which may help to generate a question that is more accurate and precise.
}

\section{Conclusion and Future work}
\label{sec:con}
In this work, 
\rev{
we have proposed a model for the task of automatic question generation based on a given code snippet.
Our model is based on sequence-to-sequence architecture, and enhanced with an \textit{attention} mechanism to perform better content selection, a \textit{copy} mechanism to handle the rare-words problem within the input code snippet as well as \textit{coverage} mechanism to discourage the meaningless repetitions.
We carried out comprehensive evaluation on Stack Overflow datasets to demonstrate the effectiveness of our approach, compared with several existing baselines, our model achieves the best performance in both the automatic evaluation and human evaluation. We have also released our code and datasets to facilitate other researchers to verify their ideas and inspire the follow up work. For future work, we plan to design better models to generate meaningful question titles by considering extra context information, such as question description. Additional work will be needed to address this context-sensitive question generation task. 
}

\section{Acknowledgements}
\label{sec:ack}

This research was partially supported by the Australian Research Council’s Discovery Early Career Researcher Award (DECRA) funding scheme (DE200100021), ARC Laureate Fellowship funding scheme (FL190100035), ARC Discovery grant DP170101932 and Singapore’s Ministry of Education (MOE2019-T2-1-193).

\bibliographystyle{ACM-Reference-Format}
\bibliography{samples}

\end{document}